\documentclass[12pt]{article}
\usepackage{epsf,epsfig,amssymb,amsmath,graphicx,float,latexsym} 

\newcommand{\lsim}{\raisebox{-0.13cm}{~\shortstack{$<$ \\[-0.07cm] $\sim$}}~}
\newcommand{\gsim}{\raisebox{-0.13cm}{~\shortstack{$>$ \\[-0.07cm] $\sim$}}~}

\begin{document}
\renewcommand{\thefootnote}{\fnsymbol{footnote}}

\begin{titlepage}
  
\begin{flushright}
April 2007
\end{flushright}

\begin{center}

\vspace{1cm}

{\Large {\bf Constraints on the Very Early Universe from Thermal WIMP Dark
    Matter}}

\vspace{1cm}

{\bf Manuel Drees}$^{a,\,}$\footnote{drees@th.physik.uni-bonn.de},
{\bf Hoernisa Iminniyaz}$^{a,b,\,}$\footnote{hoernisa@th.physik.uni-bonn.de}
and 
{\bf Mitsuru Kakizaki}$^{a,\,}$\footnote{kakizaki@th.physik.uni-bonn.de} \\

\vskip 0.15in
{\it
$^a${Physikalisches Institut der Universit\"at Bonn,
Nussallee 12, 53115 Bonn, Germany}\\
$^b${Physics Dept., Univ. of Xinjiang, 830046 Urumqi, P.R. China}\\
}
\vskip 0.5in

\abstract{We investigate the relic density $n_\chi$ of non--relativistic
  long--lived or stable particles $\chi$ in non--standard cosmological
  scenarios. We calculate the relic abundance starting from arbitrary initial
  temperatures of the radiation--dominated epoch, and derive the lower bound
  on the initial temperature $T_0 \geq m_\chi/23$, assuming that thermally
  produced $\chi$ particles account for the dark matter energy density in the
  universe; this bound holds for all $\chi$ annihilation cross sections. We
  also investigate cosmological scenarios with modified expansion rate. Even
  in this case an approximate formula similar to the standard one is capable
  of predicting the final relic abundance correctly. Choosing the $\chi$
  annihilation cross section such that the observed cold dark matter abundance
  is reproduced in standard cosmology, we constrain possible modifications of
  the expansion rate at $T \sim m_\chi/20$, well before Big Bang
  Nucleosynthesis.}

\end{center}
\end{titlepage}
\setcounter{footnote}{0}

\section{Introduction}

One of the most notable recent developments in cosmology is the precise
determination of cosmological parameters from observations of the large--scale
structure of the universe, most notably by the Wilkinson Microwave Anisotropy
Probe (WMAP). In particular, the accurate determination of the non--baryonic
cold Dark Matter (DM) density \cite{wmap},
\begin{equation} \label{e1}
0.08 < \Omega_{\rm CDM} h^2 < 0.12 \ \  (95\% \ {\rm C.L.})\, , 
\end{equation}
has great influence on particle physics models which possess dark matter
candidates \cite{kotu,dmrev}. The requirement that the predicted DM density
falls in the range (\ref{e1}) is a powerful tool for discriminating between
various models and for constraining the parameter space of surviving models.

Many dark matter candidate particles have been proposed. In particular
long--lived or stable weakly interacting massive particles (WIMPs) with
weak--scale masses are excellent candidates. In standard cosmology WIMPs
decoupled from the thermal background during the radiation--dominated epoch
after inflation.  In this framework convenient and accurate analytic
approximate solutions for the relic abundance have been derived
\cite{standard,except}. One of the best motivated candidates for WIMPs is the
lightest neutralino in supersymmetric (SUSY) models. Assuming that the
neutralino is the lightest supersymmetric particle (LSP) stabilized due to
R--parity, its relic abundance has been extensively discussed \cite{dmrev}.
Similar analyses have also been performed for other WIMPs whose existence is
postulated in other extensions of the Standard Model (SM) of particle physics.
In many cases the cosmologically favored parameter space of WIMP models can be
directly tested at the CERN Large Hadron Collider (LHC) in a few years
\cite{Ellis:2003cw}. The same parameter space often also leads to rates of
WIMP interactions with matter within the sensitivity of near--future direct DM
detection experiments.

This discussion shows that we are now entering an interesting time where the
standard cosmological scenario can be examined by experiments at high--energy
colliders as well as DM searches \cite{prebbn}. In this respect we should
emphasize that the relic abundance of thermally produced WIMPs depends not
only on their annihilation cross section, which can be determined by particle
physics experiments, but in general is also very sensitive to cosmological
parameters during the era of WIMP production and annihilation. Of particular
importance are the initial temperature $T_0$ at which WIMPs began to be
thermally produced, and the expansion rate of the universe $H$.

In the standard cosmological scenario, the expansion rate is uniquely
determined through the Friedmann equation of general relativity. In this
scenario the density of WIMPs with mass $m_\chi$ followed its equilibrium
value until the freeze--out temperature $T_F \simeq m_\chi/20$. Below $T_F$,
interactions of WIMPs are decoupled, and thus the present density is
independent of $T_0$ as long as $T_0 > T_F$.\footnote{Note that $T_F$ can be
  formally defined in the standard way even if $T_0 < T_F$. In this case WIMPs
  never were in full equilibrium, and correspondingly never ``froze out''.}

It should be noted that in non--standard scenarios the relic density can be
larger or smaller than the value in the standard scenario. One example is the
case where $T_0$ is smaller than or comparable to $T_F$, which can be realized
in inflationary models with low reheat temperature. Since in many models the
inflationary energy scale must be much higher than $m_\chi$ in order to
correctly predict the density perturbations \cite{inflatrev}, the standard
assumption $T_0 > T_F$ is not unreasonable. On the other hand, the constraint
on the reheat temperature from Big Bang Nucleosynthesis (BBN) is as low as
$T_0 \gsim {\rm MeV}$ \cite{trmin1,trmin2}. From the purely phenomenological
viewpoint, it is therefore also interesting to investigate the production of
WIMPs in low reheat temperature scenarios
\cite{trmin1,low,low_others,gg,Drees:2006vh}.

The standard scenario also assumes that entropy per comoving volume is
conserved for all temperatures $T \leq T_F$. Late entropy production can
dilute the predicted relic density \cite{late,kt}. The reason is that the usual
calculation actually predicts the {\em ratio} of the WIMP number density to
the entropy density. On the other hand, if late decays of a heavier particle
non--thermally produce WIMPs in addition to the usual thermal production
mechanism, the resulting increase of the WIMP density competes with the
dilution caused by the decay of this particle into radiation, which increases
the entropy density \cite{ckr,moroi,ad,decay,gg,endo}.

Another example of a non--standard cosmology changing the WIMP relic density
is a modified expansion rate of the universe. This might be induced by an
anisotropic expansion \cite{kt}, by a modification of general relativity
\cite{kt,Catena:2004ba}, by additional contributions to the total energy
density from quintessence \cite{Salati:2002md}, by branes in a warped geometry
\cite{Okada:2004nc}, or by a superstring dilaton \cite{Lahanas:2006xv}.

These examples show that, once the WIMP annihilation cross section is fixed,
with the help of precise measurements of the cold dark matter density we can
probe the very early stage of the universe at temperatures of ${\cal O}(m_\chi
/ 20) \sim 10$ GeV. This is reminiscent of constraining the early evolution of
the universe at $T = {\cal O}(100)~{\rm keV}$ using the primordial abundances
of the light elements produced by BBN.

The goal of this paper is to investigate to what extent the constraint
(\ref{e1}) on the WIMP relic abundance might allow us to derive quantitative
constraints on modifications of standard cosmology. So far the history of the
universe has been established by cosmological observations as far back as the
BBN era. In this paper we try to derive bounds on cosmological parameters
relevant to the era before BBN. Rather than studying specific extensions of
the standard cosmological scenario, we simply parameterize deviations from the
standard scenario, and attempt to derive constraints on these new parameters.
Since we only have the single constraint (\ref{e1}), for the most part we only
allow a single quantity to differ from its standard value. We expect that
varying two quantities simultaneously will allow to get the right relic
density for almost any WIMP annihilation cross section. This has been shown
explicitly in \cite{gg} for the case that both late entropy production and
non--thermal WIMP production are considered, even if both originate from the
late decay of a single scalar field.

We first analyze the dependence of the WIMP abundance on the initial
temperature $T_0$ of the conventional radiation--dominated epoch. We showed in
\cite{Drees:2006vh} that for fixed $T_0$ the predicted WIMP relic density
reaches a maximum as the annihilation cross section is varied from very small
to very large values. A small annihilation cross section corresponds to a
large $T_F > T_0$; in this case the relic density increases with the
annihilation cross section, since WIMP production from the thermal plasma is
more important than WIMP annihilation. On the other hand, increasing this
cross section reduces $T_F$; once $T_F < T_0$ a further increase of the cross
section leads to smaller relic densities since in this case WIMPs continue to
annihilate even after the temperature is too low for WIMP production. Here we
turn this argument around, and derive the lower bound on $T_0 \geq m_\chi/23$
under the assumption that all WIMPs are produced thermally. Note that we do
not need to know the WIMP annihilation cross section to derive this bound.

We then examine the dependence of the predicted WIMP relic density on the
expansion rate in the epoch prior to BBN, where we allow the Hubble parameter
to depart from the standard value. The standard method of calculating the
thermal relic density \cite{kotu,standard} is found to be still applicable in
this case. Our working hypothesis here is that the standard prediction for the
Hubble expansion rate is essentially correct, i.e. that the true expansion
rate differs by at most a factor of a few from the standard prediction.  We
then simply employ a generic Taylor expansion for the temperature dependence
of this modification factor; note that the success of standard BBN indicates
that this factor cannot deviate by more than $\sim 20$\% from unity at low
temperatures, $T \lsim 1$ MeV. Similarly, we assume that the WIMP annihilation
cross section has been determined (from experiments at particle colliders) to
have the value required in standard cosmology. Our approach is thus quite
different from that taken in \cite{prebbn}, where present upper bounds on the
fluxes of WIMP annihilation products are used to place upper bounds on the
Hubble expansion rate during WIMP decoupling. The advantage of their approach
is that no prior assumption on the WIMP annihilation cross section needs to be
made, whereas we assume a cross section that reproduces the correct relic
density in the standard scenario. On the other hand, the bounds derived in
refs.\cite{prebbn} are still quite weak, allowing the Hubble parameter to
exceed its standard prediction by a factor $\gsim 30$; moreover, no lower
bound on $H$ can be derived in this fashion.

The remainder of this paper is organized as follows: In Sec.~2 we will briefly
review the calculation of the WIMP relic abundance assuming a conventional
radiation--dominated universe, and derive the lower bound on the initial
temperature $T_0$. In Sec.~3 we discuss the case where the pre--BBN expansion
rate is allowed to depart from the standard one. Using approximate analytic
formulae for the predicted WIMP relic density for this scenario, we derive
constraints on the early expansion parameter.  Finally, Sec.~4 is devoted to
summary and conclusions.

\section{Relic Abundance in the Radiation--Dominated Universe}

We start the discussion of the relic density $n_\chi$ of stable or long--lived
particles $\chi$ by reviewing the structure of the Boltzmann equation which
describes their creation and annihilation. The goal of this Section is to find
the lowest possible initial temperature of the radiation--dominated universe,
assuming that the present relic abundance of cold dark matter is entirely due
to thermally produced $\chi$ particles.

As usual, we will assume that $\chi$ is self--conjugate\footnote{The case
  $\chi \neq \bar \chi$ differs in a non--trivial way only in the presence of
  a $\chi - \bar \chi$ asymmetry, i.e. if $n_\chi \neq n_{\bar \chi}$.}, $\chi
= \bar{\chi}$, and that some symmetry, for example R--parity, forbids decays
of $\chi$ into SM particles; the same symmetry then also forbids single
production of $\chi$ from the thermal background. However, the creation and
annihilation of $\chi$ pairs remains allowed. The time evolution of the number
density $n_\chi$ of particles $\chi$ in the expanding universe is then
described by the Boltzmann equation \cite{kotu},
\begin{equation} \label{eq:boltzmann_n}
  \frac{{\rm d}n_\chi}{{\rm d}t} + 3 H n_\chi =  - \langle \sigma v \rangle
  (n^2_{\chi} - n_{\chi,{\rm eq}}^2)\, ,
\end{equation}
where $n_{\chi,{\rm eq}}$ is the equilibrium number density of $\chi$, and
$\langle \sigma v \rangle$ is the thermally averaged annihilation cross
section multiplied with the relative velocity of the two annihilating $\chi$
particles. Finally, the Hubble parameter $H=\dot{R}/R$ is the expansion rate
of the universe, $R$ being the scale factor in the
Friedmann--Robertson--Walker metric. The first (second) term on the
right--hand side of Eq.(\ref{eq:boltzmann_n}) describes the decrease
(increase) of the number density due to annihilation into (production from)
lighter particles.  Eq.(\ref{eq:boltzmann_n}) assumes that $\chi$ is in
kinetic equilibrium with standard model particles.

It is useful to rewrite Eq.(\ref{eq:boltzmann_n}) in terms of the scaled
inverse temperature $x = m_\chi/T$ as well as the dimensionless quantities
$Y_\chi = n_\chi/s$ and $Y_{\chi,{\rm eq}} = n_{\chi,{\rm eq}}/s$. The entropy
density is given by $s = (2 \pi^2/45) g_{*s} T^3$, where
\begin{eqnarray}
  g_{*s} = \sum_{i = {\rm bosons}} g_i \left( \frac{T_i}{T} \right)^3
  + \frac{7}{8} \sum_{i = {\rm fermions}} g_i 
  \left( \frac{T_i}{T} \right)^3\, .
\end{eqnarray}
Here $g_i$ denotes the number of intrinsic degrees of freedom for particle
species $i$ (e.g. due to spin and color), and $T_i$ is the temperature
of species $i$. Assuming that the universe expands adiabatically, the entropy
per comoving volume, $s R^3$, remains constant, which implies
\begin{eqnarray}
  \frac{{\rm d}s}{{\rm d}t} + 3 H s = 0\, .
\end{eqnarray}
The time dependence of the temperature is then given by
\begin{eqnarray}
  \frac{{\rm d}x}{{\rm d}t} = \cfrac{Hx}
  {1 - \cfrac{x}{3g_{*s}} \cfrac{{\rm d}g_{*s}}{{\rm d}x}}\, .
\end{eqnarray}
Therefore the Boltzmann equation (\ref{eq:boltzmann_n}) can be written as
\begin{eqnarray} \label{eq:boltzmann_y}
  \frac{{\rm d}Y_\chi}{{\rm d}x} = - \frac{\langle \sigma v \rangle s}{H x}
  \left({1 - \frac{x}{3g_{*s}} \frac{{\rm d}g_{*s}}{{\rm d}x}} \right)\
  (Y_\chi^2 - Y_{\chi, {\rm eq}}^2)\, .
\end{eqnarray}

Thermal production of WIMPs takes place during the radiation--dominated epoch,
when the expansion rate is given by
\begin{eqnarray} \label{en1}
  H = \frac{\pi T^2}{M_{\rm Pl}}
  \sqrt{\frac{g_*}{90}}\, ,
\end{eqnarray}
with $M_{\rm Pl} = 2.4 \times 10^{18}$ GeV being the reduced Planck mass
and
\begin{eqnarray}
  g_{*} = \sum_{i = {\rm bosons}} g_i \left( \frac{T_i}{T} \right)^4
  + \frac{7}{8} \sum_{i = {\rm fermions}} g_i 
  \left( \frac{T_i}{T} \right)^4\, .
\end{eqnarray}
In the following we use $H_{\rm st}$ to denote the standard expansion rate
(\ref{en1}). If the post--inflationary reheat temperature was sufficiently
high, WIMPs reached full thermal equilibrium. This remains true for
temperatures well below $m_\chi$. We can therefore use the non--relativistic
expression for the $\chi$ equilibrium number density,
\begin{eqnarray}
  n_{\chi,{\rm eq}} = g_\chi \left( \frac{m_\chi T}{2 \pi} \right)^{3/2}
  {\rm e}^{-m_\chi/T}\, .
\end{eqnarray}
In the absence of non--thermal production mechanisms, $n_\chi \leq
n_{\chi,{\rm eq}}$ at early times. The annihilation rate $\Gamma = n_\chi
\langle \sigma v \rangle$ then depends exponentially on $T$, and thus drops
more rapidly with decreasing temperature than the expansion rate $H_{\rm st}$
of Eq.(\ref{en1}) does.  When the annihilation rate falls below the expansion
rate, the number density of WIMPs ceases to follow its equilibrium value and
is frozen out.

For $T \ll m_\chi$ the annihilation cross section can often (but not always
\cite{except}) be approximated by a non--relativistic expansion in powers of
$v^2$. Its thermal average is then given by
\begin{equation} \label{eq:cross_section}
\langle \sigma v \rangle = a + b \langle v^2 \rangle
  + {\cal O}(\langle v^4 \rangle) =  a + \frac{6 b}{x} 
  + {\cal O}\left( \frac{1}{x^2} \right) \, .
\end{equation}
In this standard scenario, the following approximate formula has been shown
\cite{standard,kotu,except} to accurately reproduce the exact (numerically
calculated) relic density:
\begin{eqnarray} \label{eq:relic_abundance_y}
  Y_{\chi,\infty} \equiv Y_\chi (x \to \infty)
  \simeq  \frac{1}{1.3~ m_\chi M_{\rm Pl}\sqrt{g_*(x_F)}(a/x_F + 3b/x_F^2)}\, ,
\end{eqnarray}
with $x_F = m_\chi/T_F, \ T_F$ being the decoupling temperature. For WIMPs,
$x_F \simeq 22$. Here we assume $g_* \simeq g_{*s}$ and ${\rm d}{g_*}/{\rm d}x
\simeq 0$. It is useful to express the $\chi$ mass density as $\Omega_{\chi} =
\rho_{\chi} / \rho_c$, $\rho_c = 3 H^2_0 M^2_{\rm Pl}$ being the critical
density of the universe. The present relic mass density is then given by
$\rho_{\chi} = m_\chi n_{\chi, \infty} = m_\chi s_0 Y_{\chi,\infty}$; here
$s_0 \simeq 2900~{\rm cm}^{-3}$ is the present entropy density.
Eq.(\ref{eq:relic_abundance_y}) then leads to
\begin{eqnarray} \label{eq:omegah2}
  \Omega_{\chi} h^2 = 2.7 \times 10^{10}~ Y_{\chi,\infty} 
  \left( \frac{m_\chi}{100~{\rm GeV}} \right)
  \simeq \frac{8.5 \times 10^{-11}~x_F~{\rm GeV}^{-2}}{\sqrt{g_*(x_F)}
    (a + 3 b/x_F)}\, ,
\end{eqnarray}
where $h \simeq 0.7$ is the scaled Hubble constant in units of $100$
km~sec$^{-1}$~Mpc$^{-1}$. We defer further discussions of this expression to
Sec.~3, where scenarios with modified expansion rate are analyzed. Note that
in the standard scenario leading to Eq.(\ref{eq:omegah2}), the present $\chi$
relic density is inversely proportional to its annihilation cross section and
has no dependence on the reheat temperature. Recall that this result depends
on the assumption that the highest temperature in the post--inflationary
radiation dominated epoch, which we denote by $T_0$, exceeded $T_F$
significantly.

On the other hand, if $T_0$ was too low to fully thermalize WIMPs, the final
result for $\Omega_\chi$ will depend on $T_0$. In particular, if WIMPs were
thermally produced in a completely out--of--equilibrium manner starting from
vanishing initial abundance during the radiation--dominated era, such that
WIMP annihilation remains negligible, the present relic abundance is given by
\cite{Drees:2006vh}
\begin{eqnarray} \label{eq:Y_0}
  Y_0 (x \to \infty) \simeq 0.014~ g_\chi^2 g_*^{-3/2}
  m_\chi M_{\rm Pl} {\rm e}^{-2 x_0} x_0 
  \left( a + \frac{6 b}{x_0} \right)\, .
\end{eqnarray}
Note that the final abundance depends exponentially on $T_0$, and {\em
  increases} with increasing cross section.

In in--between cases where WIMPs are not completely thermalized but WIMP
annihilation can no longer be neglected, we have shown \cite{Drees:2006vh}
that re--summing the first correction term $\delta$ enables us
to reproduce the full temperature dependence of the density of WIMPs:
\begin{eqnarray} \label{en2}
  Y_\chi \simeq \frac {Y_0} {1 - \delta/Y_0} \equiv Y_{1,r}\, .
\end{eqnarray}
Here $\delta < 0$ describes the annihilation of WIMPs produced according to
Eq.(\ref{eq:Y_0}): 
\begin{eqnarray}
  \delta(x \to \infty) \simeq - 2.5 \times 10^{-4}~
  g_\chi^4 g_*^{-5/2} m_\chi^3 M_{\rm Pl}^3
  {\rm e}^{-4 x_0} x_0 \left( a + \frac{3 b}{x_0} \right) 
  \left( a + \frac{6 b}{x_0} \right)^2\, .
\end{eqnarray}
Since $\delta$ is proportional to the third power of the cross section, the
re--summed expression $Y_{1,r}$ is inversely proportional to the cross section
for large cross section. In ref.\cite{Drees:2006vh} we have shown that this
feature allows the approximation (\ref{en2}) to be smoothly matched to the
standard result (\ref{eq:omegah2}). Not surprisingly, as long as we only
consider thermal $\chi$ production, decreasing $T_0$ can only reduce the
final $\chi$ relic density.

\begin{figure}[t!]
  \begin{center}
    \hspace*{-0.5cm} \scalebox{0.63}{\includegraphics*{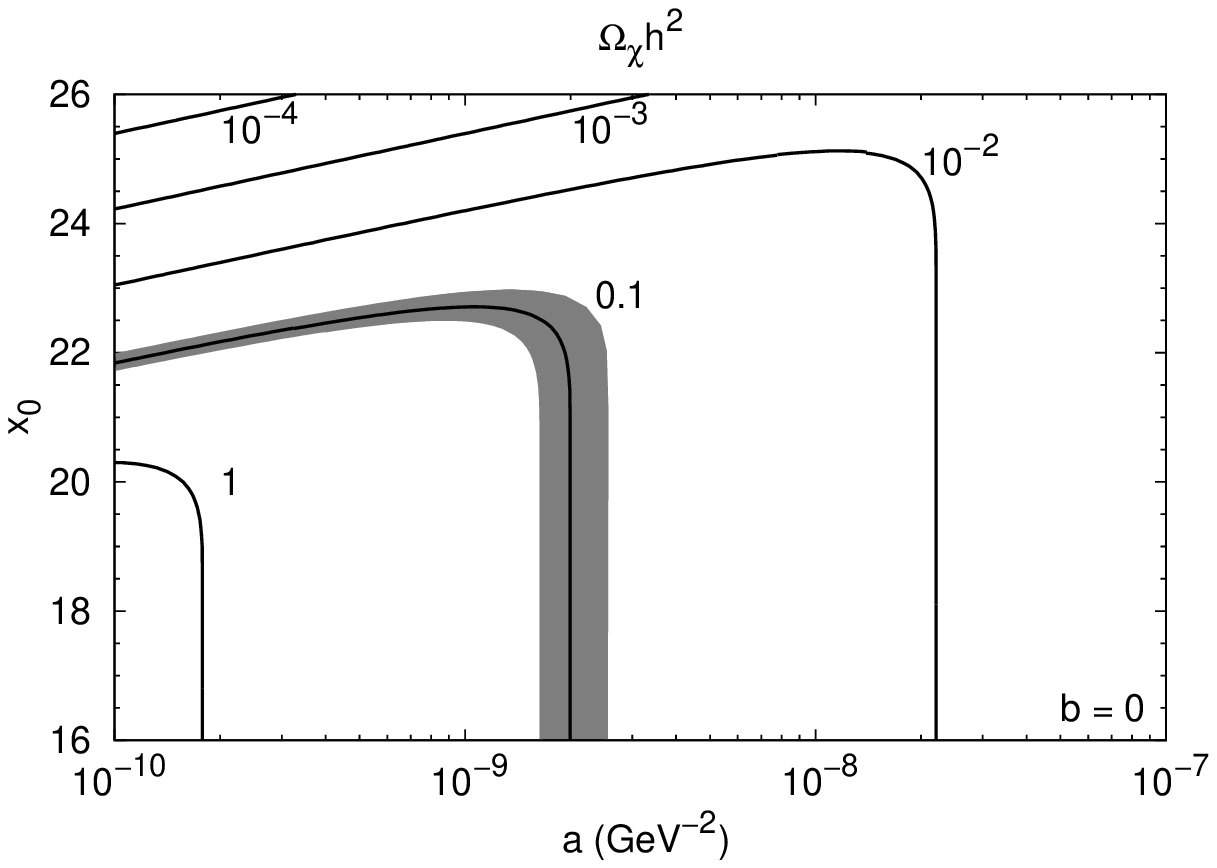}}
    \put(-115,-12){(a)} 
    \scalebox{0.63}{\includegraphics*{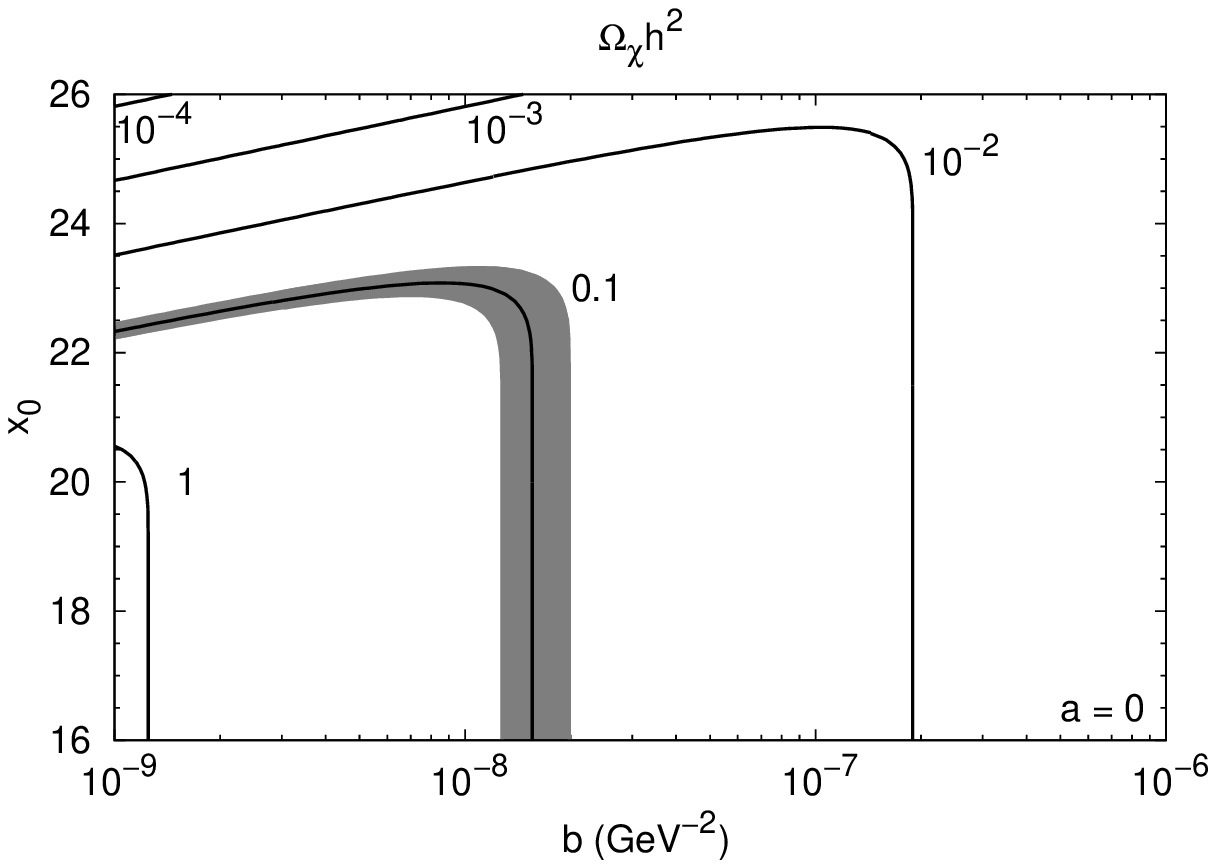}} 
    \put(-115,-12){(b)}
    \caption{\footnotesize Contour plots of the present relic abundance
      $\Omega_\chi h^2$. Here we take (a) $a \ne 0, b=0$, and (b) $a = 0, b
      \ne 0$. We choose $Y_\chi(x_0) = 0$, $m_\chi = 100$ GeV, $g_\chi = 2$,
      $g_*=90$.  The shaded region corresponds to the WMAP bound on the cold
      dark matter abundance, $0.08 < \Omega_{\rm CDM}h^2 < 0.12$ (95\% C.L.).}
    \label{fig:abundance}
  \end{center}
\end{figure}

With the help of these results, we can explore the dependence of the $\chi$
relic density on $T_0$ as well as on the annihilation cross section. Some
results are shown in Fig.~\ref{fig:abundance}, where we take (a) $a \ne 0,
b=0$, and (b) $a = 0, b \ne 0$. We choose $Y_\chi(x_0) = 0$, $m_\chi = 100$
GeV, $g_\chi = 2$ and $g_*=90$. 

The results depicted in this Figure can be understood as follows. For small
$T_0$, i.e. large $x_0$, Eq.(\ref{eq:Y_0}) is valid, leading to a very strong
dependence of $\Omega_\chi h^2$ on $x_0$. Recall that in this case the relic
density is proportional to the cross section. In this regime one can reproduce
the relic density (\ref{e1}) with quite small annihilation cross section, $a +
6 b / x_0 \lsim 10^{-9}$ GeV$^{-2}$, for some narrow range of initial
temperature, $x_0 \lsim 22.5$. Note that this allows much smaller annihilation
cross sections than the standard result, at the cost of a very strong
dependence of the final result on the initial temperature $T_0$.

In this Section we set out to derive a lower bound on $T_0$. In this regard
the region of parameter space described by Eq.(\ref{eq:Y_0}) is not optimal.
Increasing the $\chi$ annihilation cross section at first allows to obtain the
correct relic density for larger $x_0$, i.e. smaller $T_0$. However, the
correction $\delta$ then quickly increases in size; as noted earlier, once
$|\delta| > Y_0$ a further increase of the cross section will lead to a
decrease of the final relic density. The lower bound on $T_0$ is therefore
saturated if $\Omega_\chi h^2$ as a function of the cross section reaches a
maximum. From Fig.\ref{fig:abundance} we read off
\begin{equation} \label{en3}
T_0 \geq m_\chi / 23 \, ,
\end{equation}
if we require $\Omega_\chi h^2$ to fall in the range (\ref{e1}). 

We just saw that in the regime where this bound is saturated, the final relic
density is (to first order) independent of the annihilation cross section,
$\partial (\Omega_\chi h^2) / \partial \langle \sigma v \rangle = 0$. If $T_0$
is slightly above the absolute lower bound (\ref{en3}), the correct relic
density can therefore be obtained for a rather wide range of cross sections.
For example, if $x_0 = 22.5$, the entire range $3 \times 10^{-10} \ {\rm
  GeV}^{-2} \lsim a \lsim 2 \times 10^{-9} \ {\rm GeV}^{-2}$ is allowed. Of
course, the correct relic density can also be obtained in the standard
scenario of (arbitrarily) high $T_0$, if $a + 3b/22$ falls within
$\sim20\%$ of $2 \times 10^{-9} \ {\rm GeV}^{-2}$.

\section{Relic Abundance for Modified Expansion Rate}

In this section we discuss the calculation of the WIMP relic density $n_\chi$
in modified cosmological scenarios where the expansion parameter of the
pre--BBN universe differed from the standard value $H_{\rm st}$ of
Eq.(\ref{en1}). For the most part we will assume that WIMPs have been in full
thermal equilibrium. Various cosmological models predict a non--standard early
expansion history
\cite{Catena:2004ba,Salati:2002md,Okada:2004nc,Lahanas:2006xv}. Here we
analyze to what extent the relic density of WIMP Dark Matter can be used to
constrain the Hubble parameter during the epoch of WIMP decoupling. As long as
we assume large $T_0$ we can use a modification of the standard treatment
\cite{standard,kotu} to estimate the relic density for given annihilation
cross section and expansion rate. We will show that the resulting approximate
solutions again accurately reproduce the numerically evaluated relic
abundance.

Let us introduce the modification parameter $A(x)$, which parameterizes the
ratio of the standard value $H_{\rm st}(x)$ to the assumed $H(x)$:
\begin{eqnarray}
A(x) = \frac{H_{\rm st}(x)}{H(x)}\, .
\end{eqnarray}
Note that $A > 1$ means that the expansion rate is smaller than in standard
cosmology.  Allowing for this modified expansion rate, the Boltzmann equation
(\ref{eq:boltzmann_y}) is altered to
\begin{eqnarray} \label{en4}
  \frac{{\rm d}Y_\chi}{{\rm d}x} 
  = \frac{4 \pi}{\sqrt{90}} G(x) m_\chi M_{\rm Pl} 
    \frac{\langle \sigma v \rangle A(x)}{x^2} 
    \left( Y_\chi^2 - Y_{\chi, {\rm eq}}^2 \right)\, ,
\end{eqnarray}
where
\begin{eqnarray}
  G(x) = \frac{g_{*s}}{\sqrt{g_*}} \left( 1 -
  \frac{x}{3 g_{*s}} \frac{{\rm d}g_{*s}}{{\rm d}x} \right)\, .
\end{eqnarray}

Following refs.\cite{standard,kotu}, we can obtain an approximate solution of
this equation by considering the differential equation for $\Delta = Y_\chi -
Y_{\chi, {\rm eq}}$.  For temperatures higher than the decoupling temperature,
$Y_{\chi}$ tracks $Y_{\chi, {\rm eq}}$ very closely and the $\Delta^2$-term
can be ignored:
\begin{eqnarray}
  \frac{{\rm d} \Delta}{{\rm d}x} \simeq - \frac{{\rm d}Y_{\rm eq}}{{\rm d}x} 
  - \frac{4 \pi}{\sqrt{90}} m_\chi M_{\rm Pl} 
  \frac{G(x) \langle \sigma v \rangle A(x)}{x^2} 
  (2 Y_{\chi, {\rm eq}} \Delta)\, .
\end{eqnarray}
Here ${\rm d}Y_{\chi, {\rm eq}}/{\rm d}x \simeq - Y_{\chi, {\rm eq}}$ for $x
\gg 1$. In order to keep $|\Delta|$ small, the derivative ${\rm d}\Delta/{\rm
  d}x$ must also be small, which implies
\begin{eqnarray} \label{eq:high_t_solution} 
  \Delta \simeq \frac{x^2}{(8 \pi/\sqrt{90}) m_\chi M_{\rm Pl} 
    G(x) \langle \sigma v \rangle A(x)}\, .
\end{eqnarray}
This solution is used down to the freeze--out temperature $T_F$, defined via
\begin{eqnarray}
  \Delta(x_F) = \xi Y_{\chi, {\rm eq}}(x_F)\, ,
\end{eqnarray}
where $\xi$ is a constant of order of unity. This leads to the following
expression:
\begin{eqnarray} \label{en5}
  x_F = \left. \ln \left[ \sqrt{\frac{45}{\pi^5}} \xi m_\chi M_{\rm Pl} g_\chi
    \frac{\langle \sigma v \rangle A(x)}{\sqrt{x g_*(x)}} 
    \left( 1 - \frac{x}{3 g_{*s}} \frac{{\rm d}g_{*s}}{{\rm d}x} \right) 
  \right] \right|_{x=x_F}\, ,
\end{eqnarray}
which can e.g. be solved iteratively. In our numerical calculations we will
choose $\xi=\sqrt{2} -1$ \cite{standard,kotu}. 

On the other hand, for low temperatures ($T < T_F$), the production term
$\propto Y_{\chi,{\rm eq}}^2$ in Eq.(\ref{en4}) can be ignored. In this limit,
$Y_\chi \simeq \Delta$, and the solution of Eq.(\ref{en4}) is given by
\begin{eqnarray}
  \frac{1}{\Delta (x_F)} - \frac{1}{\Delta (x \to \infty)}
  = - \frac{4 \pi}{\sqrt{90}} m_\chi M_{\rm Pl} I (x_F)\, ,
\end{eqnarray}
where the annihilation integral is defined as
\begin{eqnarray} \label{eq:annihilation}
  I(x_F) = \int_{x_F}^\infty \!\! {\rm d}x \
  \frac{G(x) \langle \sigma v \rangle A(x)}{x^2}\, .
\end{eqnarray}
Assuming $\Delta (x \to \infty) \ll \Delta (x_F)$, the final relic abundance
is
\begin{eqnarray} \label{eq:relic_abundance}
  Y_{\chi,\infty} \equiv Y_\chi (x \to \infty)
 = \frac{1}{(4 \pi/\sqrt{90}) m_\chi M_{\rm Pl} I (x_F)}\, .\
\end{eqnarray}
Plugging in numerical values for the Planck mass and for today's entropy
density, the present relic density can thus be written as
\begin{eqnarray} \label{eq:relic_abundance_i}
  \Omega_\chi h^2 = \frac{8.5 \times 10^{-11}}{I(x_F)~{\rm GeV}^2}\, .
\end{eqnarray}
The constraint (\ref{e1}) therefore corresponds to the allowed range for the
annihilation integral
\begin{eqnarray}
  7.1 \times 10^{-10}~{\rm GeV}^{-2} < I(x_F) < 1.1 \times 10^{-9}~{\rm
  GeV}^{-2}\, .
\end{eqnarray}
The standard formula (\ref{eq:omegah2}) for the final relic density is
recovered if $A(x)$ is set to unity and $G(x)$ is replaced by the constant
$\sqrt{g_*(x_F)}$. 

The further discussion is simplified if we use the normalized temperature $z =
T / m_\chi \equiv 1/x$, rather than $x$. Phenomenologically $A(z)$ can be any
function subject to the condition that $A(z)$ approaches unity at late times
in order not to contradict the successful predictions of BBN. We need to know
$A(z)$ only for the interval from around the freeze-out to BBN: $z_{\rm BBN}
\sim 10^{-5} - 10^{-4} \lsim z \lsim z_F \sim 1/20$.  This suggests a
parameterization of $A(z)$ in terms of a power series in $(z - z_{F,{\rm
    st}})$:
\begin{eqnarray} \label{eq:quadratic}
  A(z) = A(z_{F,{\rm st}}) + (z - z_{F,{\rm st}}) A'(z_{F,{\rm st}})
  + \frac{1}{2} (z - z_{F,{\rm st}})^2 A''(z_{F,{\rm st}})\, ,
\end{eqnarray}
where $z_{F,{\rm st}}$ is the normalized freeze--out temperature in the
standard scenario and a prime denotes a derivative with respect to $z$. The
ansatz (\ref{eq:quadratic}) should be of quite general validity, so long as
the modification of the expansion rate is relatively modest. This suits our
purpose, since we wish to find out what constraints can be derived on the
expansion history if standard cosmology leads to the correct WIMP relic
density.

We further introduce the variable
\begin{eqnarray} \label{eq:constraint}
  k = A(z \to 0) = A(z_{F,{\rm st}}) - z_{F,{\rm st}} A'(z_{F,{\rm st}})
  + \frac{1}{2} z_{F,{\rm st}}^2 A''(z_{F,{\rm st}})\, ,
\end{eqnarray}
which describes the modification parameter at late times. Since $z_{\rm BBN}$
is almost zero, we treat $k$ as the modification parameter at the era of BBN
in this paper.\footnote{Presumably the Hubble expansion rate has to approach
  the standard rate even more closely for $T < T_{\rm BBN}$. However, since
  all WIMP annihilation effectively ceased well before the onset of BBN, this
  epoch plays no role in our analysis.}  Deviations from $k=1$ are
conveniently discussed in terms of the equivalent number of light neutrino
degrees of freedom $N_\nu$.  BBN permits that the number of neutrinos differs
from the standard model value $N_\nu=3$ by $\delta N_\nu = 1.5$ or so
\cite{bbnlimit}. We therefore take the uncertainty of $k$ to be $20\%$.  In
the following we treat $A(z_{F,{\rm st}})$, $A'(z_{F,{\rm st}})$ and $k$ as
free parameters; $A''(z_{F,{\rm st}})$ is then a derived quantity.

\begin{figure}[t!]
  \begin{center}
    \hspace*{-0.5cm} \scalebox{0.63}{\includegraphics*{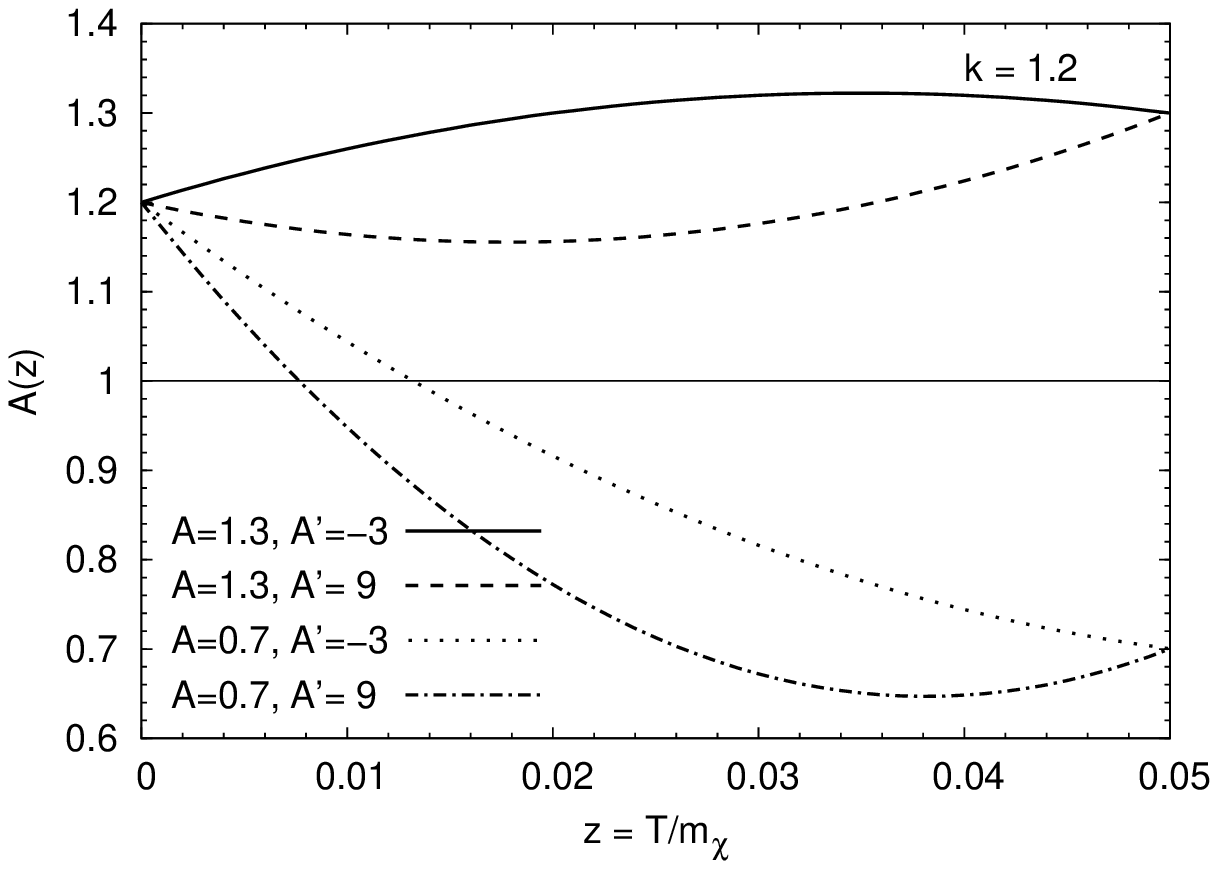}}
    \put(-115,-12){(a)} 
    \scalebox{0.63}{\includegraphics*{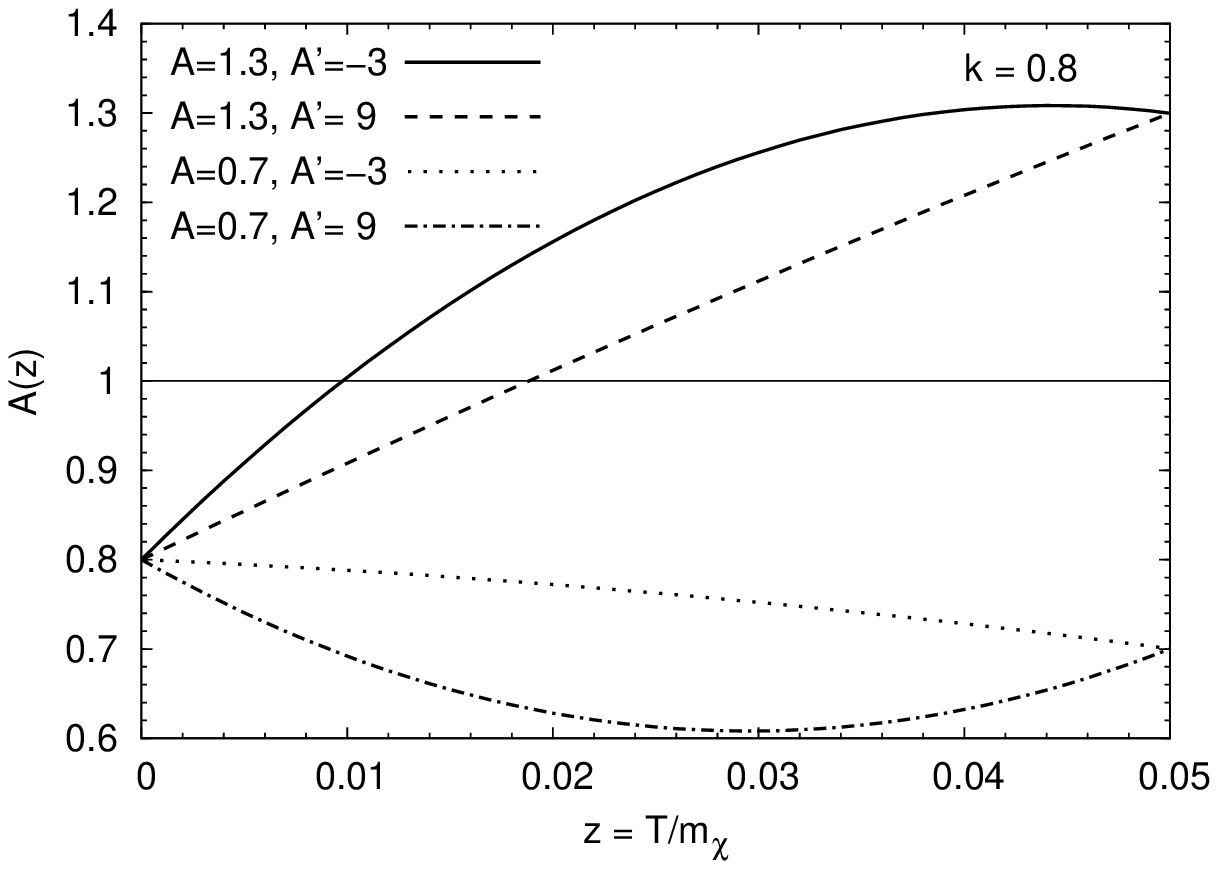}} 
    \put(-115,-12){(b)}
    \caption{\footnotesize 
      Examples of possible evolutions of the modification parameter $A(z)$ as
      function of $z$ for $z_F=0.05$.  Here we take $k=1.2$ (left frame) and
      $k=0.8$ (right).  In each frame we choose $A(z_F) = 1.3, A'(z_F) = - 3$
      (thick line), $A(z_F) = 1.3, A'(z_F) = 9$ (dashed), $A(z_F) = 0.7,
      A'(z_F) = - 3$ (dotted), $A(z_F) = 0.7, A'(z_F) = 9$ (dot--dashed).}
    \label{fig:a}
  \end{center}
\end{figure}

Note that we allow $A(z)$ to cross unity, i.e. to switch from an expansion
that is faster than in standard cosmology to a slower expansion or vice versa.
This is illustrated in Fig.~\ref{fig:a}, which shows examples of possible
evolutions of $A(z)$ as function of $z$ for $z_F = 0.05$. Here we take $k=1.2$
(left frame) and $k=0.8$ (right). In each case we consider scenarios with
$A(z_F) = 1.3$ (slower expansion at $T_F$ than in standard cosmology) as well
as $A(z_F) = 0.7$ (faster expansion); moreover, we allow the change of $A$ at
$z=z_F$ to be either positive or negative. However, we insist that $H$ remains
positive at all times, i.e. $A(z)$ must not cross zero. This excludes
scenarios with very large positive $A'(z_{F,{\rm st}})$, which would lead to
$A < 0$ at some $z < z_F$.  Similarly, demanding that our ansatz
(\ref{eq:quadratic}) remains valid for some range of temperatures above $T_F$
excludes scenarios with very large negative $A'(z_{F,{\rm st}})$. We will come
back to this point shortly.

\begin{figure}[t!]
  \begin{center}
    \scalebox{0.63}{\includegraphics*{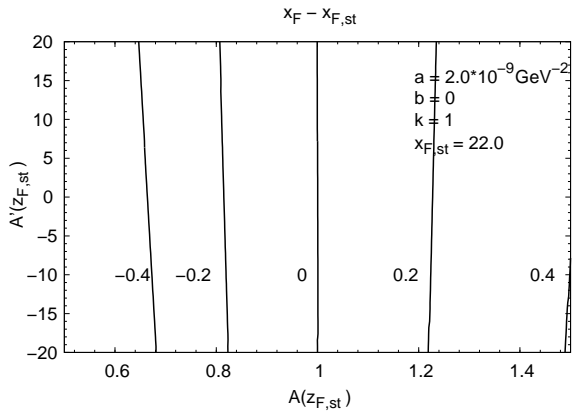}}
    \caption{\footnotesize 
      Contour plot of $x_F - x_{F,{\rm st}}$ in the ($A(z_{F,{\rm st}})$,
      $A'(z_{F,{\rm st}})$) plane.  Here we take $a = 2.0 \times 10^{-9}$
      GeV$^{-2}$, $b=0$, $m_\chi = 100$ GeV, $g_\chi = 2$, $g_*=90$ (constant)
      and $k=1$. This parameter set produces $x_{F,{\rm st}} = 22.0$ and
      $\Omega_\chi h^2 = 0.099$ for the standard approximation.}
    \label{fig:deltaxf}
  \end{center}
\end{figure}

Eq.(\ref{en5}) shows that $z_F \ne z_{F,{\rm st}}$ ($x_F \ne x_{F,{\rm st}}$)
if $A(z_F) \neq 1$. This is illustrated by Fig.~\ref{fig:deltaxf}, which shows
the difference between $x_F$ and $x_{F,{\rm st}}$ in the ($A(z_{F,{\rm st}})$,
$A'(z_{F,{\rm st}})$) plane. Here we take parameters such that $\Omega_\chi
h^2 = 0.099$ in the standard cosmology, which is recovered for $A(z_{F,{\rm
    st}})=1$, $A'(z_{F,{\rm st}})=0$. Due to the logarithmic dependence on
$A$, $x_F$ (or $z_F$) differs by at most a few percent from its standard value
if $A(z_{F,{\rm st}})$ is ${\cal O}(1)$. Since $T_F$ only depends on the
expansion rate at $T_F$, it is essentially insensitive to the derivative
$A'(z_{F,{\rm st}})$.

In our treatment the modification of the expansion parameter affects the WIMP
relic density mostly via the annihilation integral (\ref{eq:annihilation}). In
terms of the normalized temperature $z$, the latter can be rewritten as
\begin{eqnarray}
  I(z_F) = \int_0^{z_F} \!\! {\rm d}z\ G(z) \langle \sigma v \rangle A(z)\, .
\end{eqnarray}
One advantage of the expansion (\ref{eq:quadratic}) is that this integral can
be evaluated analytically:
\begin{eqnarray} \label{eq:iofzf}
  I(z_F) & \simeq & G(z_F) \left[
    k (a z_F + 3b z_F^2)
    +  (A'(z_{F,{\rm st}}) - z_{F,{\rm st}} A''(z_{F, {\rm st}})) 
    \left( \frac{a}{2} z_F^2 + 2b z_F^3 \right) \right. 
  \nonumber \\
  && \qquad \left. + \frac{A''(z_{F,{\rm st}})}{2} 
    \left( \frac{a}{3} z_F^3 + \frac{3b}{2} z_F^4 \right) \right]\, .
\end{eqnarray}
Here we have assumed that $G(z)$ varies only slowly. 

\begin{figure}[t!]
  \begin{center}
    \hspace*{-0.5cm} 
    \scalebox{0.63}{\includegraphics*{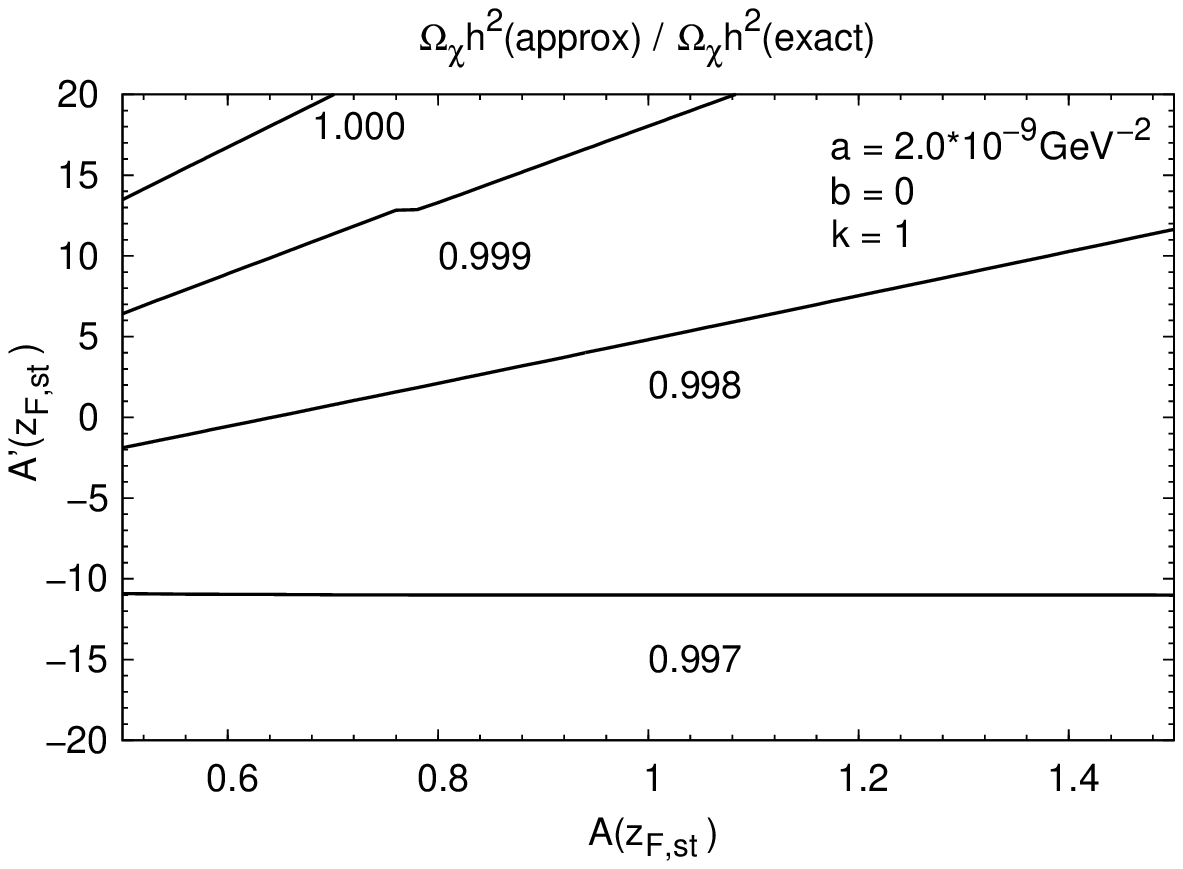}}
    \put(-115,-12){(a)} 
    \scalebox{0.63}{\includegraphics*{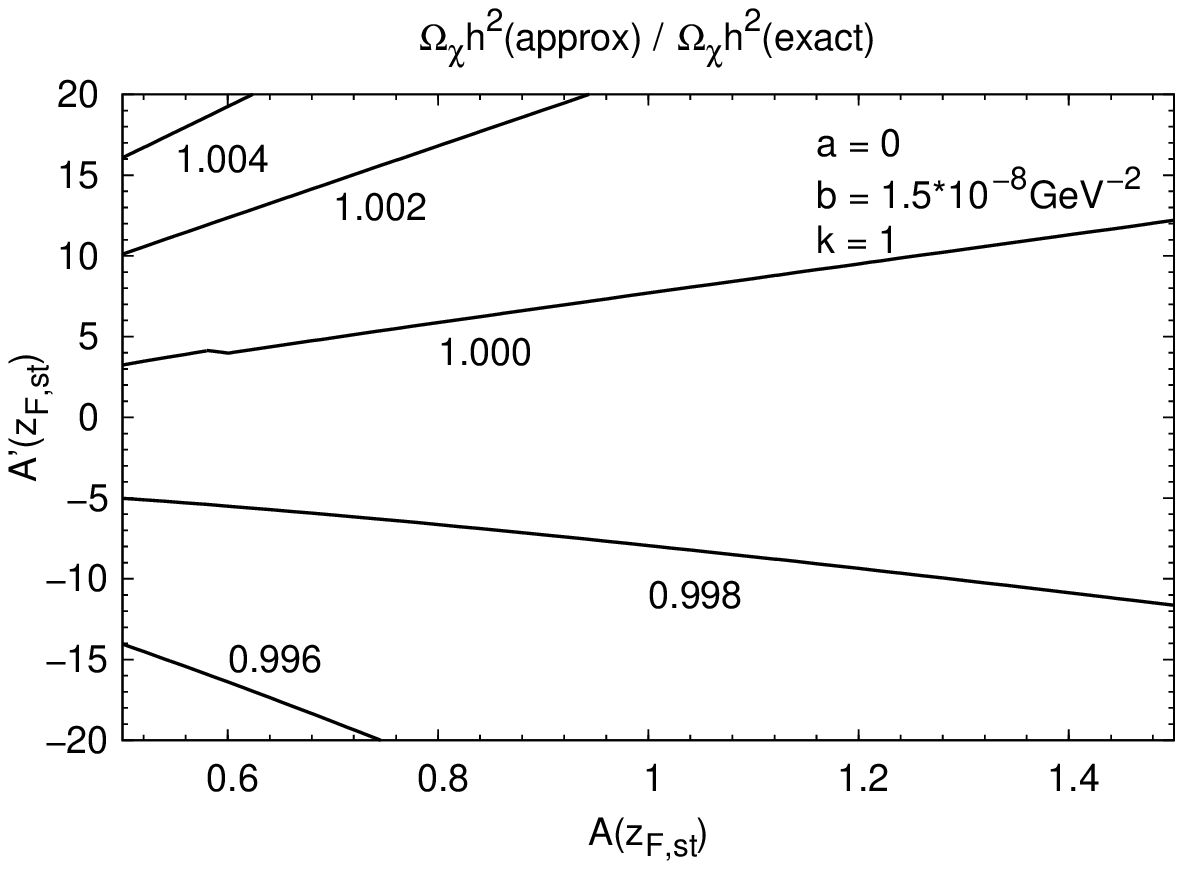}}
    \put(-115,-12){(b)} 
    \caption{\footnotesize 
      Ratio of the analytic result of the relic density to the exact value in
      the ($A(z_{F,st})$, $A'(z_{F,st})$) plane for $a = 2.0 \times 10^{-9}$
      GeV$^{-2}$, $b=0$ (left frame) and for $a=0$, $b = 1.5 \times 10^{-8}$
      GeV$^{-2}$ (right). The other parameters are as in
      Fig.~\ref{fig:deltaxf}.}
    \label{fig:ratio}
  \end{center}
\end{figure}

Before proceeding, we first have to convince ourselves that the analytic
treatment developed in this Section still works for $A \neq 1$. This is
demonstrated by Fig.~\ref{fig:ratio}, which shows the ratio of the analytic
solution obtained from Eqs. (\ref{eq:relic_abundance_i}) and (\ref{eq:iofzf})
to the exact one, obtained by numerically integrating the Boltzmann equation
(\ref{en4}), assuming constant $g_*$. We see that our analytical treatment is
accurate to better than 1\%, and can thus safely be employed in the subsequent
analysis.
  
\begin{figure}[t!]
  \begin{center}
    \hspace*{-0.5cm} \scalebox{0.63}{\includegraphics*{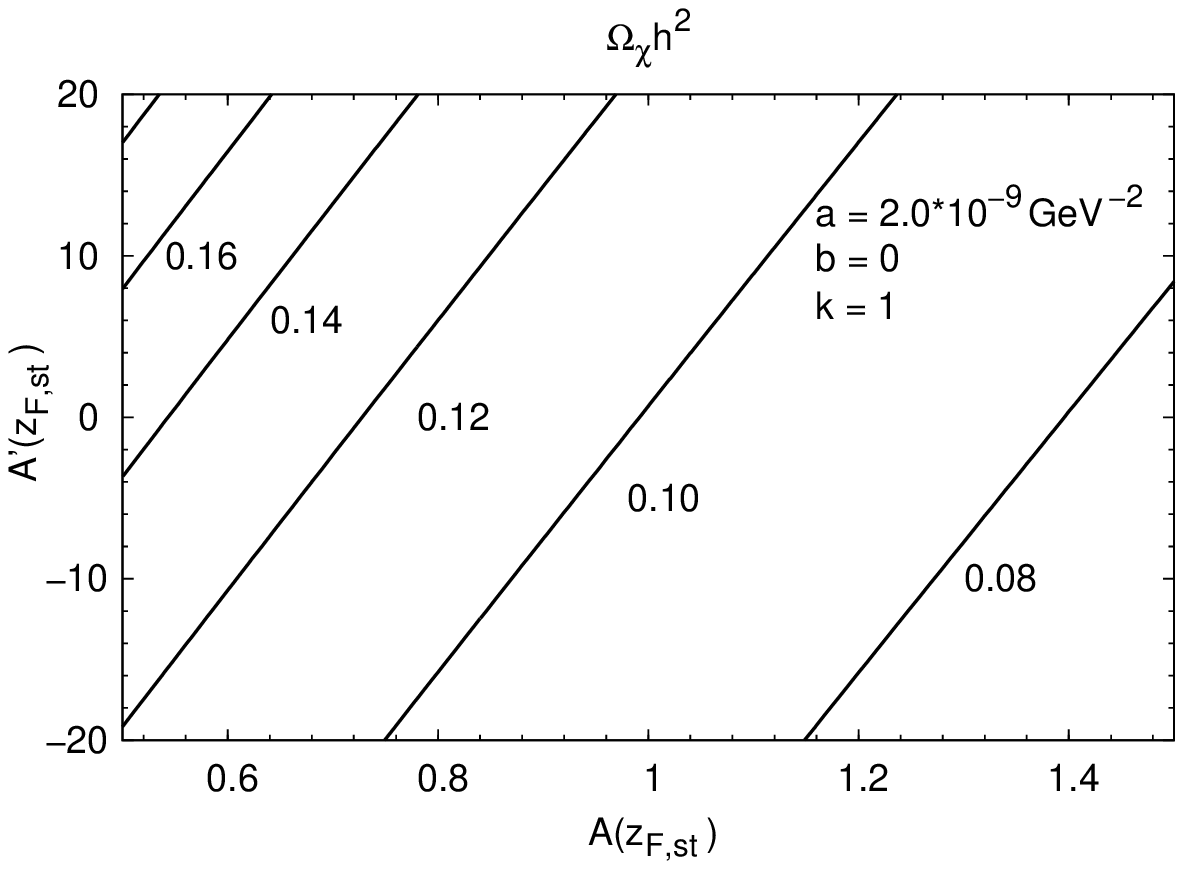}}
    \put(-115,-12){(a)} 
    \scalebox{0.63}{\includegraphics*{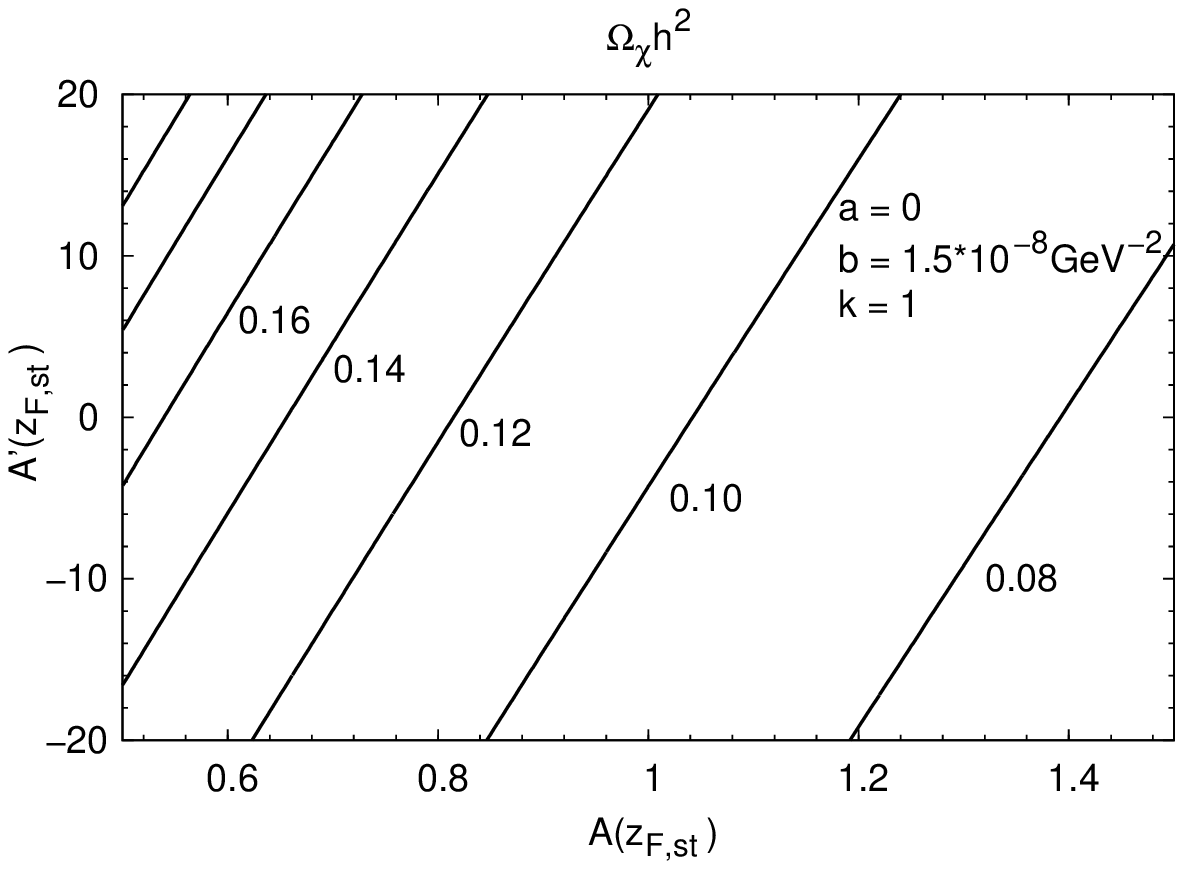}} 
    \put(-115,-12){(b)} 
    \vspace{0.5cm}
    \hspace*{-0.5cm} \scalebox{0.63}{\includegraphics*{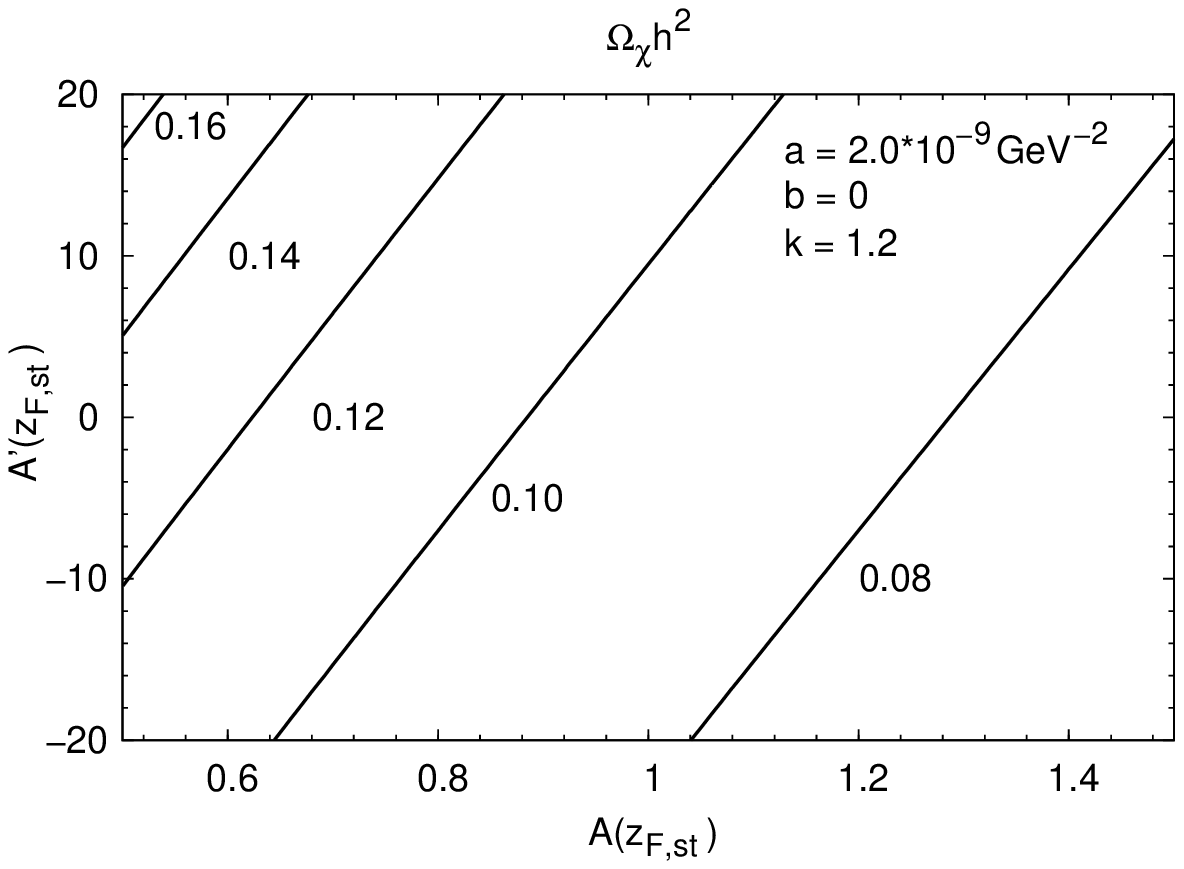}}
    \put(-115,-12){(c)} 
    \scalebox{0.63}{\includegraphics*{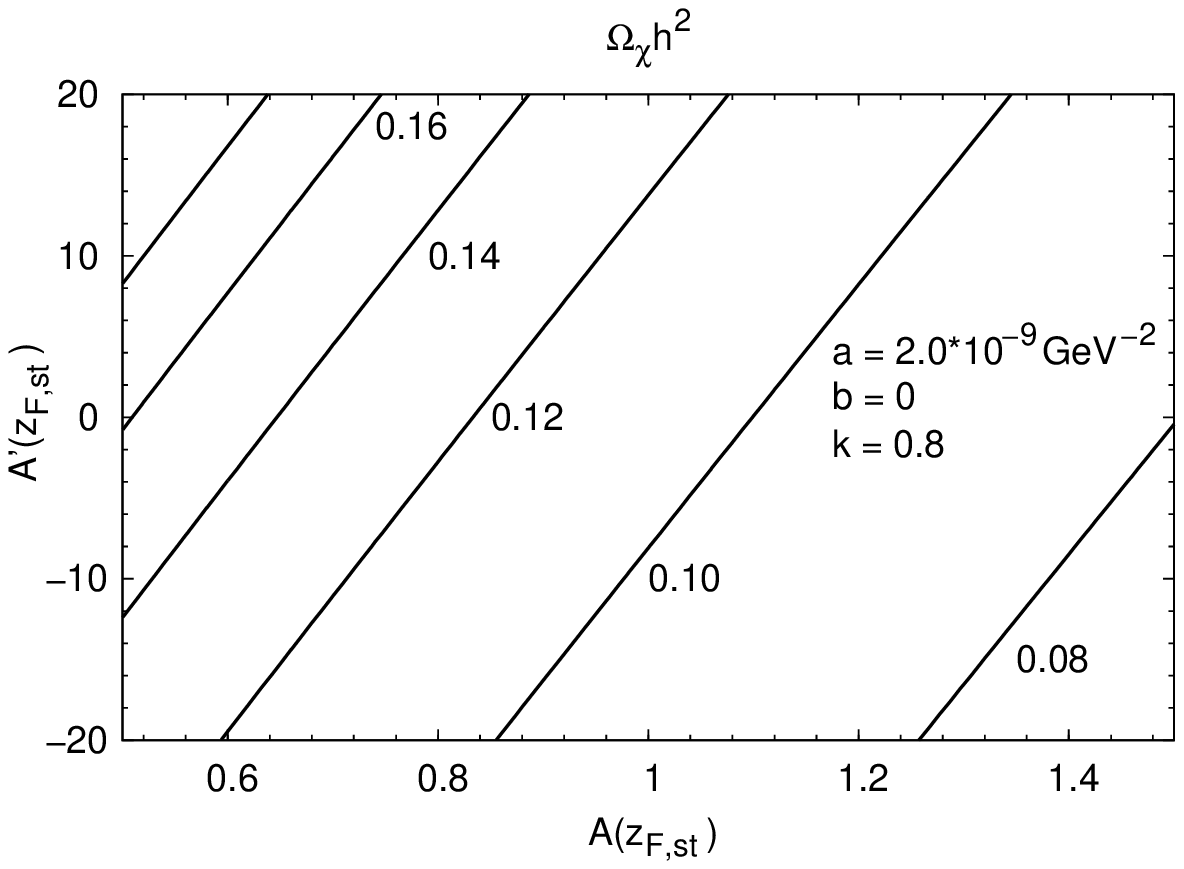}} 
    \put(-115,-12){(d)} 
    \caption{\footnotesize 
      Contour plots of the relic abundance in the ($A(z_{F,st})$,
      $A'(z_{F,st})$) plane. Here we choose 
      (a) $a = 2.0 \times 10^{-9}$ GeV$^{-2}$, $b=0$, $k=1$;
      (b) $a=0$, $b = 1.5 \times 10^{-8}$ GeV$^{-2}$, $k=1$;
      (c) $a = 2.0 \times 10^{-9}$ GeV$^{-2}$, $b=0$, $k=1.2$;
      (d) $a = 2.0 \times 10^{-9}$ GeV$^{-2}$, $b=0$, $k=0.8$.
      The other parameters are as in Fig.~\ref{fig:deltaxf}.}
    \label{fig:contour}
  \end{center}
\end{figure}

We are now ready to analyze the impact of the modified expansion rate on the
WIMP relic density.  In Fig.~\ref{fig:contour}, we show contour plots of
$\Omega_\chi h^2$ in the ($A(z_{F,{\rm st}})$, $A'(z_{F,{\rm st}})$) plane.
Recall that large (small) values of $A$ correspond to a small (large)
expansion rate. Since a smaller expansion rate allows the WIMPs more time to
annihilate, $A > 1$ leads to a reduced WIMP relic density, whereas $A < 1$
means larger relic density, if the cross section is kept fixed.

However, unlike the freeze--out temperature, the annihilation integral is
sensitive to $A(z)$ for all $z \leq z_F$. Note that $A'(z_{F,{\rm st}}) > 0$
implies $A(z) < A(z_{F,{\rm st}})$ for $z < z_{F,{\rm st}} \simeq z_F$. A
positive first derivative, $A'(z_{F,{\rm st}}) > 0$, can therefore to some
extent compensate for $A(z_{F,{\rm st}}) > 1$; analogously, a negative first
derivative can compensate for $A(z_{F,{\rm st}}) < 1$. This explains the
slopes of the curves in Fig.~\ref{fig:contour}. Recall also that $A'(z_{F,{\rm
    st}}) = 0$ does not imply a constant modification factor $A(z)$; rather,
the term $\propto A''(z_{F,{\rm st}})$ in Eq.(\ref{eq:quadratic}) makes sure
that $A$ approaches $k$ as $z \rightarrow 0$. This explains why a change of
$A$ by some given percentage leads to a {\em smaller} relative change of
$\Omega_\chi h^2$, as can be seen in the Figure. This also illustrates the
importance of ensuring appropriate (near--standard) expansion rate in the BBN
era. Finally, since the expansion rate at late times is given by $H_{\rm
  st}/k$, bigger (smaller) values of $k$ imply that the $\chi$ relic density
is reduced (enhanced).

Fig.~\ref{fig:contour} shows that we need additional physical constraints if
we want to derive bounds on $A(z_{F,{\rm st}})$ and/or $A'(z_{F,{\rm st}})$.
Once the annihilation cross section is known, the requirement (\ref{e1}) will
single out a region in the space spanned by our three new parameters
(including $k$) which describe the non--standard evolution of the universe,
but this region is not bounded. Such additional constraints can be derived
from the requirement that the Hubble parameter should remain positive
throughout the epoch we are considering.  As noted earlier, requiring $H > 0$
for all $T < T_{F,{\rm st}}$ leads to an upper bound on $A'(z_{F,{\rm st}})$;
explicitly,
\begin{eqnarray} \label{en6}
  A'(z_{F,{\rm st}})  < \frac{2 \left(A(z_{F,{\rm st}}) + \sqrt{k A(z_{F,{\rm 
  st}})}\right)} {z_{F,{\rm st}}}\, .
\end{eqnarray}
On the other hand, a lower bound on $A'(z_{F,{\rm st}})$ is obtained from the
condition that the modified Hubble parameter is positive between the highest
temperature $T_i$ where the ansatz (\ref{eq:quadratic}) holds and $T_{F,\rm
  st}$:
\begin{eqnarray} \label{en7}
  A'(z_{F,{\rm st}}) > - \left[ \frac{1}{z_i - z_{F,{\rm st}}}
    \left( 2 - \frac{z_i}{z_{F,{\rm st}}} \right)
    A(z_{F,{\rm st}}) + k \left( \frac{1}{z_{F,{\rm st}}} 
      - \frac{1}{z_i} \right) \right]\, ,
\end{eqnarray}
for $(1 - z_{F,{\rm st}}/z_i)^2 k < A(z_{F,{\rm st}})$, and 
\begin{eqnarray} \label{en8}
  A'(z_{F,{\rm st}}) 
  > \frac{2 \left( A(z_{F,{\rm st}}) - \sqrt{k A(z_{F,{\rm st}})} \right)}
  {z_{F,{\rm st}}}\, ,
\end{eqnarray}
for $A(z_{F,{\rm st}}) < (1 - z_{F,{\rm st}}/z_i)^2 k$, where $z_i =
T_i/m_\chi$.  

\begin{figure}[t!]
  \begin{center}
    \hspace*{-0.5cm} \scalebox{0.63}{\includegraphics*{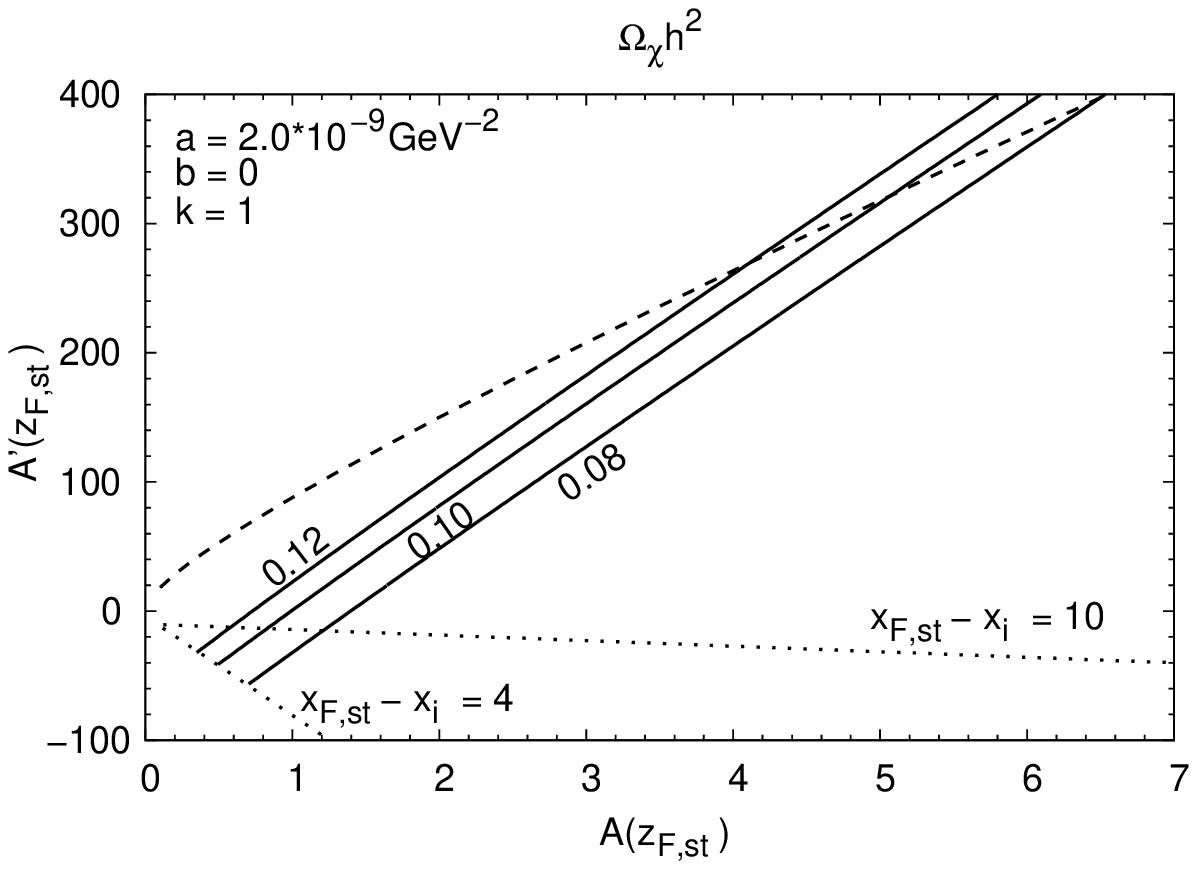}}
    \put(-115,-12){(a)} 
    \scalebox{0.63}{\includegraphics*{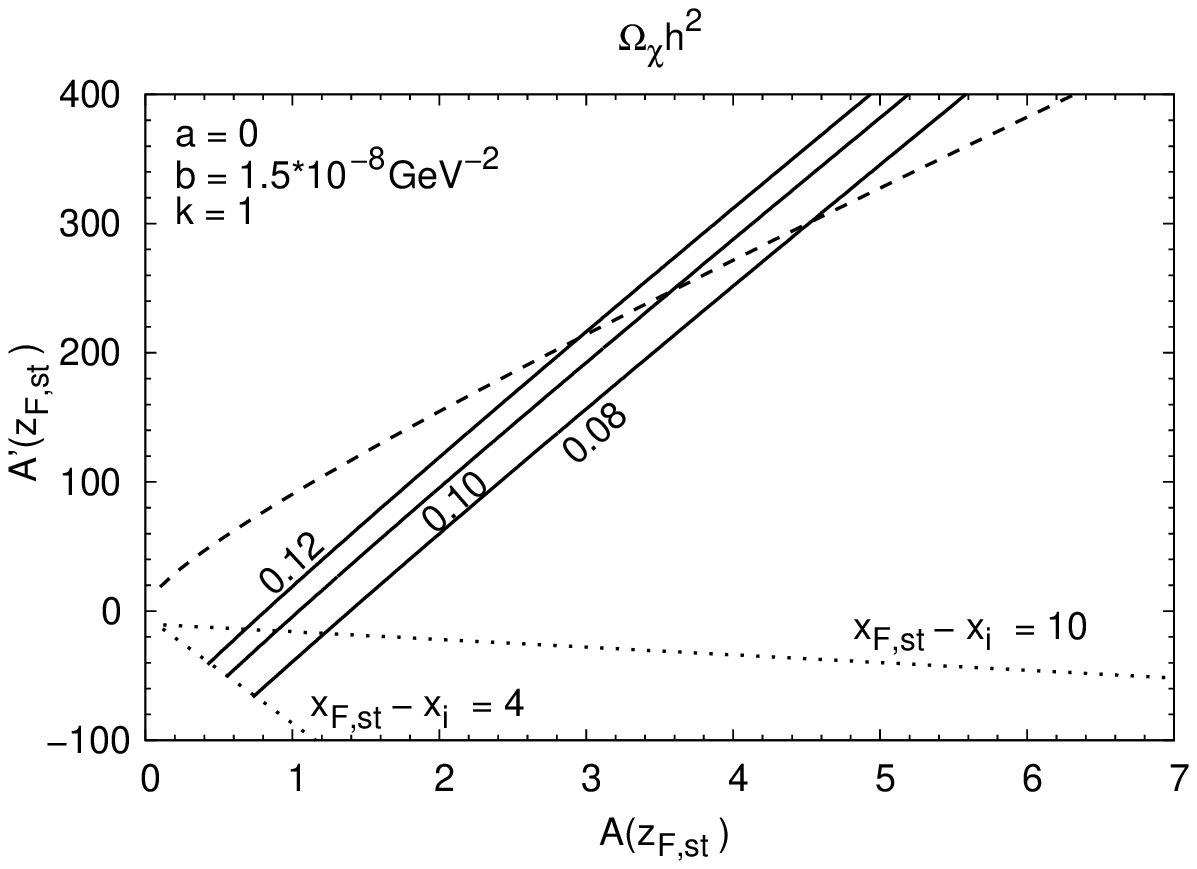}}
    \put(-115,-12){(b)}
     \caption{\footnotesize 
       Contour plots of the relic abundance $\Omega_\chi h^2$ in the
       $(A(z_{F,{\rm st}}), A'(z_{F,{\rm st}}))$ plane.  The dashed line
       corresponds to the upper bound on $A'(z_{F,{\rm st}})$.  The dotted
       lines correspond to the lower bounds calculated for $x_{F,{\rm st}} -
       x_i = 4, 10$. We take $a = 2.0 \times 10^{-9}$ GeV$^{-2}, \, b = 0$
       (left frame) and $a = 0, \, b = 1.5 \times 10^{-8}$ GeV$^{-2}$ (right
       frame).  The other parameters are as in Fig~\ref{fig:deltaxf}.}
    \label{fig:constraints}
  \end{center}
\end{figure}

Evidently the lower bound on $A'(z_{F,{\rm st}})$ depends on $z_i$, i.e. on
the maximal temperature where we assume our ansatz (\ref{eq:quadratic}) to be
valid. In ref.\cite{Drees:2006vh} we have shown that in standard cosmology ($A
\equiv 1$) essentially full thermalization is already achieved for $x_i \lsim
x_F - 0.5$, even if $n_\chi(x_i) = 0$. However, it seems reasonable to demand
that $H$ should remain positive at least up to $x_i = x_F -$(a few). In
Fig.~\ref{fig:constraints} we therefore show the physical constraints on the
modification parameter $A(z)$ for $x_{F,{\rm st}} - x_i = 4, 10$ and $k=1$.
The dashed and dotted lines correspond to the upper and lower bounds on
$A'(z_{F,{\rm st}})$, described by Eq.(\ref{en6}) and Eqs.(\ref{en7}),
(\ref{en8}), respectively. We see that when $x_{F,{\rm st}} - x_i = 4$ the
allowed region is $0.4 \lsim A(z_{F,{\rm st}}) \lsim 6.5$ with $- 60 \lsim
A'(z_{F,{\rm st}}) \lsim 400$ for $b = 0$ (left frame), and $0.4 \lsim
A(z_{F,{\rm st}}) \lsim 4.5$ with $- 60 \lsim A'(z_{F,{\rm st}}) \lsim 300$
for $a=0$ (right frame).  When $x_{F,{\rm st}} - x_i = 10$, the lower bounds
are altered to $0.6 \lsim A(z_{F,{\rm st}})$, $- 10 \lsim A'(z_{F,{\rm st}})$
for $b=0$ (left frame), and $0.6 \lsim A(z_{F,{\rm st}})$, $-20 \lsim
A'(z_{F,{\rm st}})$ for $a=0$ (right frame).  Note that the lower bounds on
$A(z_{F,{\rm st}})$, which depend only weakly on $x_i$ so long as it is not
very close to $x_F$, are almost the same in both cases, which also lead to
very similar relic densities in standard cosmology.  However, the two upper
bounds differ significantly.  The reason is that large values of $A(z_{F,{\rm
    st}})$, i.e. a strongly suppressed Hubble expansion, require some degree
of finetuning: One also has to take large positive $A'(z_{F,{\rm st}})$, so
that $A$ becomes smaller than one for some range of $z$ values below $z_F$,
leading to an annihilation integral of similar size as in standard cosmology.
Since the $b-$terms show different $z_F$ dependence in the annihilation
integral (\ref{eq:iofzf}), the required tuning between $A(z_{F,{\rm st}})$ and
$A'(z_{F,{\rm st}})$ is somewhat different than for the $a-$terms, leading to
a steeper slope of the allowed region. This allowed region therefore saturates
the upper bound (\ref{en6}) on the slope for somewhat smaller $A(z_{F,{\rm
    st}})$.

\begin{figure}[t!]
  \begin{center}
    \hspace*{-0.5cm} \scalebox{0.63}{\includegraphics*{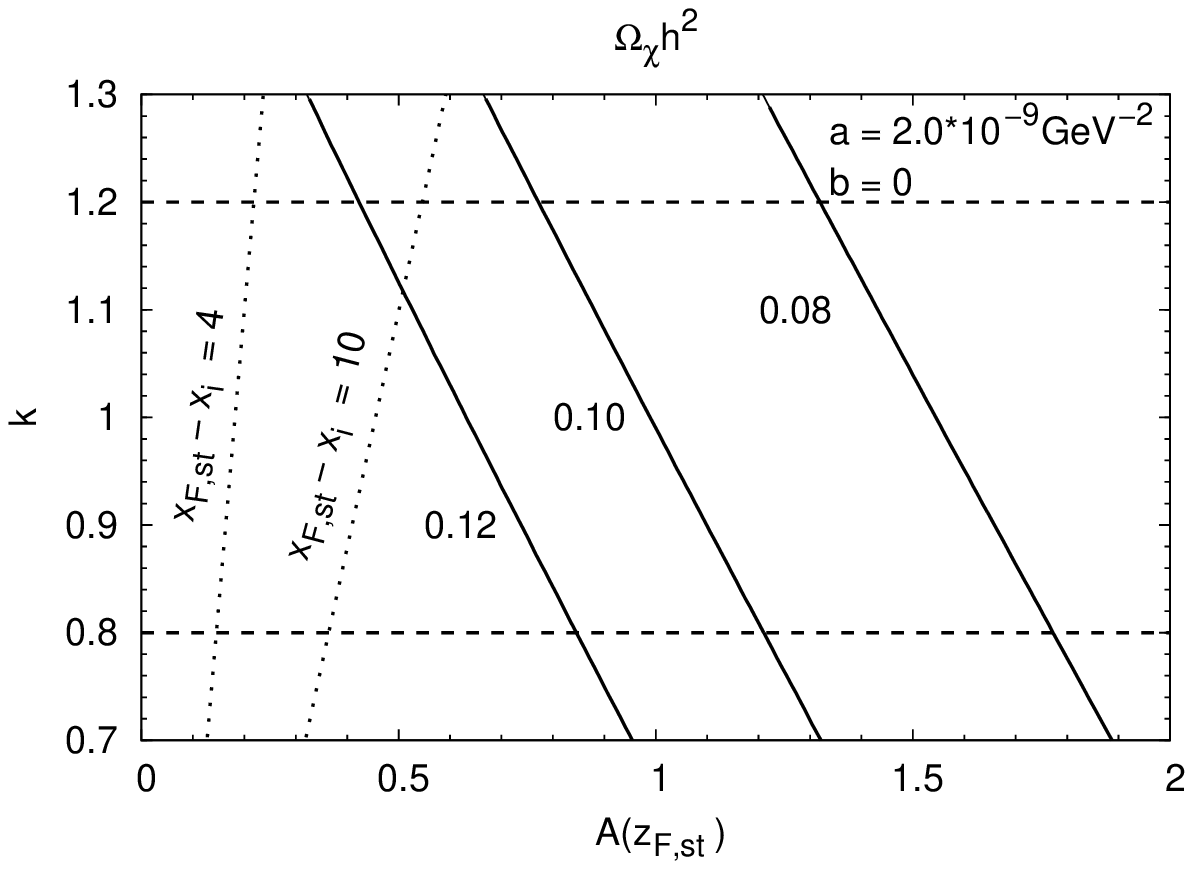}}
    \put(-115,-12){(a)} 
    \scalebox{0.63}{\includegraphics*{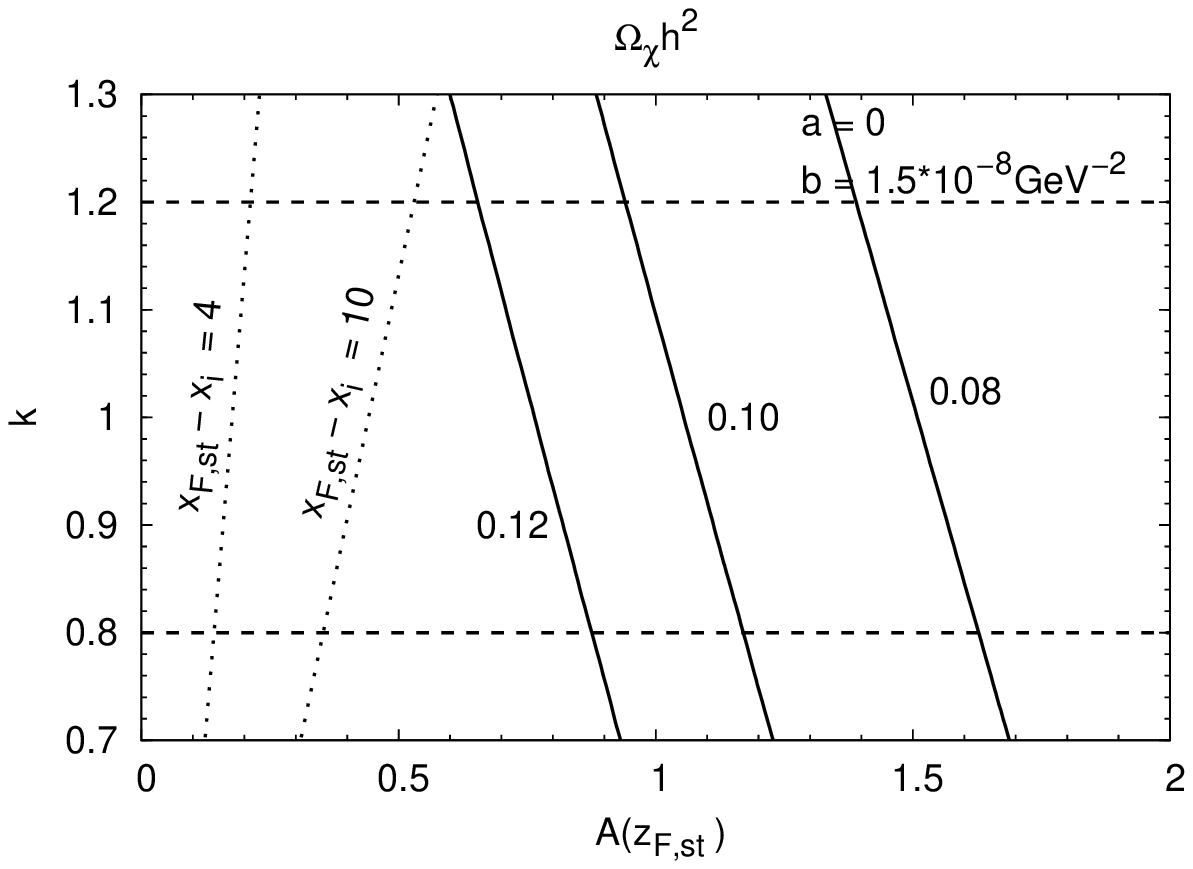}}
    \put(-115,-12){(b)}
     \caption{\footnotesize 
       Contour plots of the relic abundance $\Omega_\chi h^2$ in the
       $(A(z_{F,{\rm st}}), k)$ plane for $A''(z_{F,{\rm st}}) = 0$. The
       dotted lines correspond to the lower bounds of $A(z_{F,{\rm st}})$,
       calculated for $x_{F,{\rm st}} - x_i = 4, 10$. We take $a = 2.0 \times
       10^{-9}$ GeV$^{-2}, \, b = 0$ (left frame) and $a = 0, \, b = 1.5
       \times 10^{-8}$ GeV$^{-2}$ (right frame). The other parameters are as
       in Fig~\ref{fig:deltaxf}.}
    \label{fig:new}
  \end{center}
\end{figure}

The effect of this tuning can be seen by analyzing the special case where
$A''(z_{F,{\rm st}})=0$. The modification parameter then reads
\begin{eqnarray}
   A(z) = \frac{A(z_{F,{\rm st}}) - k}{z_{F,{\rm st}}} z + k\, .
\end{eqnarray}
Note that $A$ is now a monotonic function of $z$, making large cancellations
in the annihilation integral impossible. Imposing that $A(z)$ remains positive
for $z_{F,{\rm st}} \leq z \leq z_i$ leads to the lower limit
\begin{eqnarray} \label{en9}
  A(z_{F,{\rm st}}) > \left( 1 - \frac{z_{F,{\rm st}}}{z_i} \right) k\, .
\end{eqnarray}
There is no upper bound, since $A(z)$ is now automatically positive for all $z
\in [0,z_{F,{\rm st}}]$ if $A(z_{F,{\rm st}})$ and $A(0) \equiv k$ are both
positive.  Fig.~\ref{fig:new} shows constraints on the relic abundance in the
$(A(z_{F,{\rm st}}),k)$ plane for $A''(z_{F,{\rm st}}) = 0$.  The dotted lines
correspond to the lower bounds (\ref{en9}) on $A(z_{F,{\rm st}})$ for
$x_{F,{\rm st}} - x_i = 4, 10$. As noted earlier, $k$ is constrained by the
BBN bound. This leads to the bounds $0.5 \lsim A(z_{F,{\rm st}}) \lsim 1.8$
for $b = 0$ (left frame), and $0.65 \lsim A(z_{F,{\rm st}}) \lsim 1.6$ for $a
= 0$ (right frame), when $x_{F,{\rm st}} - x_i = 10$. Evidently the
constraints now only depend weakly on whether the $a-$ or $b-$term dominates
in the annihilation cross section. As the initial temperature is lowered, the
impact of the constraint (\ref{en9}) disappears.

So far we have assumed in this Section that the reheat temperature was high
enough for WIMPs to have attained full thermal equilibrium.  If this was not
the case, the initial temperature as well as the suppression parameter affects
the final relic abundance. Here we show that the lower bound on the reheat
temperature derived in the previous Section survives even in scenarios with
altered expansion history as long as WIMPs were only produced thermally.

This can be understood from the observation that the Boltzmann equation with
modified expansion rate is obtained by replacing $\langle \sigma v \rangle $
in the radiation--dominated case by $\langle \sigma v \rangle A$. Increasing
(decreasing) $A$ therefore has the same effect as an increase (decrease) of the
annihilation cross section. Since the lower bound on $T_0$ was independent of
$\sigma$ (more exactly: we quoted the absolute minimum, for the optimal choice
of $\sigma$), we expect it to survive even if $A(z) \neq 1$ is introduced.

\begin{figure}[t!]
  \begin{center}
    \scalebox{0.63}{\includegraphics*{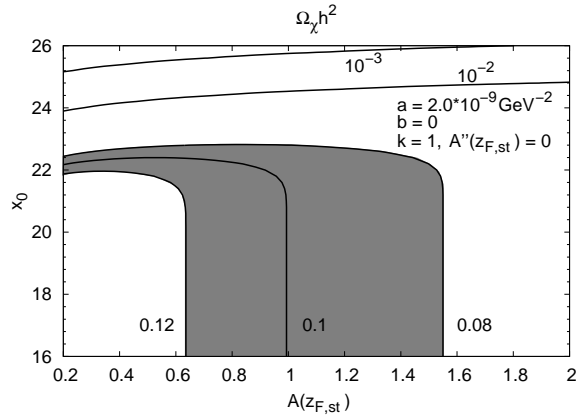}}
    \caption{\footnotesize 
      Contour plot of the relic abundance $\Omega_\chi h^2$ in the
      ($A(z_{F,st})$, $x_0$) plane. Here we choose $a = 2.0 \times 10^{-9}$
      GeV$^{-2}$, $b=0$, $k=1, A''(z_{F,{\rm st}})=0$.  The other parameters
      are as in Fig.~\ref{fig:deltaxf}.  The shaded region corresponds to
      the WMAP bound on the cold dark matter abundance, $0.08 < \Omega_{\rm
        CDM}h^2 < 0.12$ (95\% C.L.).}
    \label{fig:oh2_a-x0}
  \end{center}
\end{figure}

This is borne out by Fig.~\ref{fig:oh2_a-x0}, which shows the relic abundance
$\Omega_\chi h^2$ in the ($A(z_{F,st})$, $x_0$) plane for the simplified case
$A''(z_{F,{\rm st}}) = 0$; similar results can be obtained for the more
general ansatz (\ref{eq:quadratic}). The shaded region corresponds to the bound
(\ref{e1}) on the cold dark matter abundance. As expected, this figure looks
similar to Fig.~\ref{fig:abundance} if the annihilation cross section in
Fig.~\ref{fig:abundance} is replaced by $A(z_{F,{\rm st}})$. The
maximal value of $x_0$ consistent with the WMAP data remains around $23$ even
in these scenarios with modified expansion rate.
Fig.~\ref{fig:oh2_a-x0} also shows that $A(z_{F,{\rm st}}) \ll 1$  is allowed
for some narrow range of initial temperature $T_0 < T_F$.  This is analogous
to the low cross section branch in Fig.~\ref{fig:abundance}.

\section{Summary and Conclusions}

In this paper we have investigated the relic abundance of WIMPs $\chi$, which
are nonrelativistic long--lived or stable particles, in non--standard
cosmological scenarios. One motivation for studying such scenarios is that
they allow to reproduce the observed Dark Matter density for a large range of
WIMP annihilation cross sections. Our motivation was the opposite: we wanted
to quantify the constraints that can be obtained on parameters describing the
early universe, under the assumption that thermally produced WIMPs form all
Dark Matter. Wherever necessary, we fixed particle physics quantities such
that standard cosmology yields the correct relic density.

Specifically, we first considered scenarios with low post--inflationary reheat
temperature, while keeping all other features of standard cosmology (known
particle content and Hubble expansion parameter during WIMP decoupling; no
late entropy production; no non--thermal WIMP production channels). If the
temperature was so low that WIMPs could not achieve full thermal equilibrium,
the dependence of the abundance on the mass and annihilation cross section of
the WIMPs is completely different from that in the standard thermal WIMP
scenario. In particular, if the maximal temperature $T_0$ is much less than
the decoupling temperature $T_F$, $n_\chi$ remains exponentially suppressed.
By applying the observed cosmological amount of cold dark matter to the
predicted WIMP abundance, we therefore found the lower bound of the initial
temperature $T_0 \gsim m_\chi/23$. One might naively think that this bound
could be evaded by choosing a sufficiently large WIMP production (or
annihilation) cross section. However, increasing this cross section also
reduces $T_F$. For sufficiently large cross section one therefore has $T_F
\leq T_0$ again; in this regime the relic density drops with increasing cross
section. Our lower bound is the minimal $T_0$ required for {\em any} cross
section; once the latter is known, the bound on $T_0$ might be slightly
sharpened. As a by--product, we also noted that the final relic density
depends only weakly on the annihilation cross section if $T_0$ is slightly
above this lower bound.

We also investigated the effect of a non--standard expansion rate of the
universe on the WIMP relic abundance. In general the abundance of thermal
relics depends on the ratio of the annihilation cross section to the expansion
rate; the latter is determined unambiguously in standard cosmology. We found
that even for non--standard Hubble parameter the relic abundance can be
calculated accurately in terms of an annihilation integral, very similar to
the case of standard cosmology. We assumed that the WIMP annihilation cross
section is such that the standard scenario yields the observed relic density,
and parameterized the modification of the Hubble parameter as a quadratic
function of the temperature. In this analysis it is crucial to make sure that
at low temperatures the Hubble parameter approaches its standard value to
within $\sim 20\%$, as required for the success of Big Bang Nucleosynthesis
(BBN). 

Keeping the annihilation cross section fixed and allowing a 20\% variation in
the relic density, roughly corresponding to the present ``2$\sigma$'' band, we
found that the expansion of the universe at $T=T_F$ might have been more than
two times faster, or more than six times slower, than in standard cosmology.
These large variations of $H(T_F)$ can only be realized by finetuning of the
parameters describing $H(T < T_F)$. However, even if we forbid such finetuning
by choosing a linear parameterization for the modification of the expansion
rate, a $20\%$ variation of $\Omega_\chi h^2$ allows a difference between
$H(T_F)$ and its standard expectation of more than 50\%. This relatively weak
sensitivity of the predicted $\Omega_\chi h^2$ on $H(T_F)$ is due to the fact
that the relic density depends on {\em all} $H(T < T_F)$; as stressed above,
we have to require that $H(T \ll T_F)$ approaches its standard value to within
$\sim 20\%$. The fact that determining $\Omega_\chi h^2$ will yield relatively
poor bounds on $H(T_F)$ remains true even if the annihilation cross section is
such that a non--standard behavior of $H(T)$ is required for obtaining the
correct $\chi$ relic density. Finally, we showed that the absolute lower bound
on the temperature required for thermal $\chi$ production is unaltered by
allowing $H(T)$ to differ from its standard value.

Of course, in order to draw the conclusions derived in this article, we need
to convince ourselves that WIMPs do indeed form (nearly) all Dark Matter. This
requires not only the detection of WIMPs, e.g. in direct search experiments;
we also need to show that their density is in accord with the local Dark
Matter density derived from astronomical observations. To that end, the cross
sections appearing in the calculation of the detection rate need to be known
independently. This can only be done in the framework of a definite theory,
using data from collider experiments. For example, in order to determine the
cross section for the direct detection of supersymmetric WIMPs, one needs to
know the parameters of the supersymmetric neutralino, Higgs and squark sectors
\cite{dmrev}. We also saw that inferences about $H(T_F)$ can only be made if
the WIMP annihilation cross section is known. This again requires highly
non--trivial analyses of collider data \cite{coll}, as well as a consistent
overall theory. We thus see that the interplay of accurate cosmological data
with results obtained from dark matter detections and collider experiments can
give us insight into the pre--BBN universe, which to date remains unexplored
territory.

\subsection*{Acknowledgments}
This work was partially supported by the Marie Curie Training Research Network
``UniverseNet'' under contract no.  MRTN-CT-2006-035863, as well as by the
European Network of Theoretical Astroparticle Physics ENTApP ILIAS/N6 under
contract no.  RII3-CT-2004-506222.

\end{document}